\documentclass[a4paper,11pt]{article}
\pdfoutput=1 

\usepackage{jheppub} 

\usepackage{hyperref}
\usepackage{fancybox}
\usepackage{float}
\usepackage{amsfonts}
\usepackage{amsmath}
\usepackage{amsbsy}
\usepackage{graphicx}
\usepackage{amssymb}
\usepackage{amsthm}
\usepackage{bm}
\usepackage{epsfig}
\usepackage{latexsym}
\usepackage{pdflscape}
\usepackage{color}

\makeatletter
\def\@fpheader{\relax}
\makeatother

\DeclareSymbolFont{AMSa}{U}{msa}{m}{n}
\DeclareSymbolFont{AMSb}{U}{msb}{m}{n}
\DeclareMathSymbol{\fieldR}{\mathalpha}{AMSb}{"52}

\DeclareMathOperator{\tr}{tr}
\DeclareMathOperator{\pf}{pf}

\DeclareMathOperator{\sgn}{sgn}
\DeclareMathOperator{\sn}{sn}
\DeclareMathOperator{\cn}{cn}
\DeclareMathOperator{\dn}{dn}

\DeclareMathOperator{\Det}{Det}

\newcommand{\beq}{\begin{eqnarray}}
\newcommand{\eeq}{\end{eqnarray}}
\newcommand{\bea}{\begin{eqnarray}}
\newcommand{\eea}{\end{eqnarray}}
\newcommand{\be}{\begin{equation}}
\newcommand{\ee}{\end{equation}}
\newcommand{\bq}{\begin{equation}}
\newcommand{\eq}{\end{equation}}
\newcommand{\unit}{1\!\!1}
\newcommand{\half}{\frac{1}{2}}
\newcommand{\nn}{\nonumber}
\def\tb{\tilde{\beta}}

\def\lab{\label}

\def\t{\tau}
\def\e{\epsilon}
\def\o{\omega}
\def\le{\left}
\def\ri{\right}

\def\half{\frac12}
\def\q{\theta}

\def\6{\partial}

\def\a{\alpha}

\def\lab{\label}

\def\g{\gamma}
\def\tr{{\rm Tr}}

\def\lab{\label}
\def\le{\left}
\def\ri{\right}
\def\6{\partial}

\title{\boldmath Matrix Quantum Mechanics on $S^{1}/{\mathbb Z}_{2}$}

\author{P. Betzios,}
\author{U. G\"ursoy}
\author{and O. Papadoulaki}
\affiliation{Institute for Theoretical Physics and Center for Extreme Matter and Emergent Phenomena,\\
Utrecht University,\\ Princetonplein 5, 3584 CC Utrecht, The Netherlands.}

\emailAdd{P.Betzios@uu.nl}
\emailAdd{U.Gursoy@uu.nl}
\emailAdd{O.Papadoulaki@uu.nl}

\abstract{We study Matrix Quantum Mechanics on the Euclidean time orbifold $S_1/\mathbb{Z}_2$. Upon Wick rotation to Lorentzian time and taking the double-scaling limit this theory provides a toy model for a big-bang/big crunch universe in two dimensional non-critical string theory where the orbifold fixed points become cosmological singularities. We derive the MQM partition function both in the canonical and grand canonical ensemble in two different formulations  and demonstrate agreement between them. We pinpoint the contribution of twisted states in both of these formulations  either in terms of bi-local operators acting at the end-points of time or branch-cuts on the complex plane. We calculate, in the matrix model, the contribution of the twisted states to the torus level partition function explicitly and show that it precisely matches the world-sheet result, providing a non-trivial test of the proposed duality. Finally we discuss some interesting features of the partition function and the possibility of realising it as a $\tau$-function of an integrable hierarchy.}
\arxivnumber{1612.04792}

\begin{document} 
\maketitle
\flushbottom

\section{Introduction and Motivation}
\label{sec:intro}

Little is known in Quantum Gravity about how the space-like singularities in general, and cosmological singularities in particular, can be resolved---if they can be resolved at all. Some of the questions in this context are: is string theory able to provide consistent, non-singular dynamics around such singularities? What is the set of possible initial conditions for the cosmological evolution starting from a big-bang singularity? What are the possible initial wave-functions of the universe at the big-bang? How is the evolution of the universe determined following the big-bang, et cetera. 
Quantum gravity is, notoriously, a subject where problems vastly outnumber results especially for the physics near spacetime singularities. At short distances strong fluctuations of the metric are expected to cause a breakdown of classical geometry and the notion of space and time might lose their meaning and become emergent concepts of a more fundamental theory. Nevertheless, various efforts to understand the initial conditions of the Universe based on the semi-classical approximation to the path integral were made in the 80's, most notably the no-boundary proposal of Hartle and Hawking~\cite{Hartle:1983ai} and the tunnelling boundary condition of Linde and Vilenkin~\cite{Vilenkin:1982de,Linde:1983mx}  but it is fair to say we do not have a unique sensible answer to the aforementioned problems.

It is natural to ponder what string theory has to offer in this context, and whether it can resolve these problems or at least provide a new perspective. In string/M-theory these fundamental questions have been addressed in the various approximations and using various models in the past. Some notable work includes the study of time dependent orbifolds and the null-brane construction~\cite{Liu:2002kb,Liu:2002ft,Balasubramanian:2002ry,Fabinger:2002kr,Robbins:2005ua,Martinec:2006ak}, the Bang-Crunch scenarios~\cite{Elitzur:2002rt,Craps:2003ai,Turok:2004gb,Freedman:2004xg,Craps:2005wd,Craps:2006xq,Ishino:2006nx,Blau:2008bp}, tachyon condensation~\cite{McGreevy:2005ci,Hikida:2005xa,Nakayama:2006gt}, constructions attempting to address cosmological singularities via string theory~\cite{Antoniadis:1988vi,Larsen:1996pb,Cornalba:2002nv,Craps:2002ii,Berkooz:2002je,Florakis:2010is,Krishnan:2013cra}, or via AdS/CFT~\cite{Turok:2007ry,Craps:2007ch,Craps:2008cj,Engelhardt:2015gta,Brandenberger:2016egn,Kumar:2015jxa} and pre-Big Bang scenarios \cite{Gasperini:2002bn} among many others. For related work on string cosmology with a view towards inflation see~\cite{Quevedo:2002xw,Kachru:2003aw,McAllister:2007bg}.

Motivated by these difficult questions, we ask a more modest question in this paper: What can string theory teach us about the cosmological singularities in the context of a toy model: the two dimensional non-critical string theory or $c=1$ Liouville theory\footnote{More precisely $c=1$ Liouville theory is an exact CFT equivalent to 2D string theory in a linear dilaton and exponential tachyon background in the Liouville direction $\phi$. The non-critical is an adjective referring to the number of dimensions.}~\cite{Polyakov:1981rd,Nakayama:2004vk}? The idea here is the following\footnote{The basic idea and some of the calculations presented in this paper are due to discussions that one of the authors (U.G.) had together with Hong Liu in 2005 \cite{GL}.}. Start with the {\em Euclidean} 2D non-critical string theory with Euclidean time direction, $\tau$, compactified on a circle with radius $R$. This theory has a well-known dual formulation in terms of Matrix Quantum Mechanics (MQM) of a Hermitean $N \times N$ dimensional matrix $M$ at finite temperature $T = 1/2\pi R$ in a double scaling limit~\cite{Gross:1990ub,Klebanov:1991qa,Kazakov:1990ue}. Now, consider a $\mathbb{Z}_2$ orbifold of the non-critical string theory (NCST) in the Euclidean time direction where one identifies $ \tau \sim -\tau$. The following identifications
\be\lab{ids} 
\tau \sim \tau + 2\pi R, \qquad \textrm{and}\qquad \tau \sim - \tau\, ,
\ee
restrict the domain of the Euclidean time to the line segment $0 \leq \tau \leq \pi R$.
Upon Wick rotating to Lorentzian time, the fixed points of the orbifold at the points $\tau =0$ and $\tau = \pi R$ correspond respectively to the big-bang and big-crunch singularities of a toy, cosmological big-bang/big crunch universe in two dimensions. The questions posed above are expected to have a much simpler formulation in this toy universe, since the only non-trivial physical degrees of freedom in the bulk are a massless closed string ``tachyon field''  in case of the bosonic NCST with an additional RR scalar $C_0$ in case of supersymmetric type 0B NCST~\cite{Polyakov:1981re,Kutasov:1990ua,DiFrancesco:1991daf,Douglas:2003up,Takayanagi:2003sm}. This is to be contrasted with the infinitely many physical excitations of the critical bosonic string in 26 dimensions and supersymmetric string in 10 dimensions. The 2D toy model also enjoys the following great advantage: Resolution of cosmological singularities in string theory is expected to involve not only the full set of corrections in the string length scale $\alpha'$ but also the perturbative corrections in the string coupling constant $g_s$~\cite{Kutasov:1990sv}\footnote{and possibly corrections non-perturbative in $g_s$.}. This seems an insurmountable task for critical string theories (unless one attempts to use the BFSS~\cite{Banks:1996vh} or related matrix model formulations, as in some of the references above). In the case of 2D NCST however, the dual formulation in terms of Hermitean MQM comes to the rescue. The partition function evaluated via MQM involves at least the full set of perturbative $g_s$ corrections in the dual string theory and in addition a lot is understood for the non-perturbative corrections as well~\cite{Alexandrov:2003nn,Martinec:2003ka,Kazakov:2004du,Alexandrov:2004cg}. 

The duality between 2D NCST and the Hermitean MQM was discovered in late 80s~\cite{Kazakov:1985ea,Boulatov:1986jd,Kazakov:1988ch,Brezin:1989ss}. Starting from a Lagrangian of the form 
\be\lab{lagr} 
{\cal L}  = \tr\le( \frac12 \le(\frac{\6 M}{\6 t} \ri)^2 + \frac{1}{2\alpha'} M^2 - \frac{\kappa}{ 3!} M^3\ri)\, ,
\ee 
where $M$ is a Hermitean N by N matrix, one constructs the web of Feynman diagrams that arise from the cubic interaction vertex. This web of Feynman diagrams then provides the dual lattice of the one obtained from triangulations of a string world-sheet a la 't Hooft~\cite{tHooft:1973alw}. As one increases the bare coupling $\kappa$ one discovers that the average number of triangles on a given world-sheet begins to diverge at a critical value $\kappa_c$. Then, taking the double scaling limit $N\to \infty$, $\kappa\to \kappa_c$ with $N(\kappa_c^2 - \kappa^2)$ kept constant, one obtains a continuum formulation of the 2D string theory in terms of matrix quantum mechanics. The crucial point here is that a {\em universality} arises in this double scaling limit, that focuses on the tip of the potential provided by the mass term in (\ref{lagr}). Therefore, the theory dual to the continuum limit of the 2D string theory is just described by Hermitean matrix quantum mechanics  with the {\em inverse harmonic oscillator potential}.
In this duality, the time direction in MQM provides the time direction for the 2D space-time where the string can propagate. In addition, the eigenvalues $\lambda_i$ of the matrix $M$ provide the extra space-like Liouville direction $\phi$ in the 2D string theory picture. 
 
In some sense this duality is the oldest example of the open/close dualities in string theory, much before the famous AdS/CFT correspondence in the critical IIB string theory \cite{Maldacena:1997re}. The lessons learned from AdS/CFT, in particular the role of D-branes in this correspondence, ignited a revival of interest in the old matrix quantum mechanics in the 00s. A gauge/gravity type of interpretation focusing on the target space physics arising from the matrix model has been proposed in~\cite{McGreevy:2003kb,Klebanov:2003km}. According to this picture, MQM describes the field theory living on N $D0$ branes, the ZZ branes found in~\cite{Zamolodchikov:2001ah}, that sit at the strong coupling end of the Liouville theory. Furthermore, the $0B$ fermionic NCST also admits a non-perturbative formulation where the cubic potential in case of the bosonic NCST is simply replaced by a quartic potential. Therefore, unlike the bosonic theory, $0B$ fermionic NCST is believed to be non-perturbatively stable \cite{Douglas:2003up,Takayanagi:2003sm}. One important insight that arises from the D-brane interpretation in the string/matrix duality is the need to introduce a non-dynamical\footnote{This gauge field is necessarily non-dynamical in two dimensions.} bulk gauge field $A_0(\tau)$ in the matrix path integral. Integration over this gauge field then projects to the singlet sector of the MQM. The gauged matrix model then captures the physics of the so-called linear dilaton background of the 2D string theory.

In this paper, we consider a toy cosmological universe with a big bang/big crunch singularity
in the context of bosonic and $0B$ NCST. As explained above, a natural model that is suitable for this purpose is a space-time where the (Euclidean) time direction is compactified and orbifolded as $S^{1}/\mathbb{Z}_{2}$ and coupled to the Liouville direction. If the Euclidean time direction in this model admits
an analytic continuation into Lorentzian signature, one can interpret the orbifold singularities
as cosmological singularities. One also hopes that information about the initial and
final wavefunctions is encoded in the twisted sector of the orbifold that describes states localized at the orbifold fixed points. One can go further and also ask if one can compute the transition
amplitude of the universe in this model. 

We take the first step toward this aim in this paper and focus on the calculation of the orbifold matrix model partition function in {\em Euclidean} time using the machinery of matrix quantum mechanics. In particular, we show that
\begin{itemize} 
\item the orbifold operation is represented in the matrix model by the operation $\textrm{diag}(-1,-1,\\ \cdots -1, 1,1, \cdots, 1) \,\,\star$ where $\star$ acts on time as $\star t = -t \star$ and with $n$   eigenvalues with the value $-1$ in the diagonal matrix. Hence, there are $n=0, \cdots N/2$ distinct orbifold representations on the matrix model. In the T-dual D-instanton picture there are $N-2n$ fractional instantons that are stuck at the orbifold fixed points and $n$ D-particles free to move along the t-directon. We argue that the correct choice corresponds to $n=N/2$, where there are no fractional instantons.
\item Using the matrix model techniques we calculate the torus partition function in the large $R$ limit for the $n=0$ and $n=N/2$ representations, especially the twisted state contribution to it, and show that the $n=N/2$ representation matches precisely the result obtained from the world-sheet CFT. This provides a non-trivial check of the equivalence we propose between the orbifold MQM and the orbifold 2D non-critical string theory. 
\item The calculation of the full orbifold partition function in the canonical ensemble in the large-N limit proves hard. However, we manage to represent the grand-canonical partition function in terms of an integral kernel whose spectrum gives the single-particle density of states. We obtain this density by two independent methods that agree with each other. 
\item We further discuss certain aspects of this matrix model in connection with the corresponding 2D string theory. Finally we make various comments on how to implement the Wick rotation of the Euclidean time orbifold partition function to Lorentzian signature. We leave the full Lorentzian space-time interpretation of the possible initial and final boundary conditions at the cosmological singularities and a more thorough study of the semi-classical geometry that the matrix model describes, to future work. 
\end{itemize} 
The organization of the paper is as follows. In the next section we first outline the necessary material on the orbifold $c=1$ Liouville theory. In particular we present the torus partition function of the bosonic, super-affine $0B$ and $0A$ NCSTs including the contribution from the twisted sectors. This is achieved by considering the possible $\mathbb{Z}_2$ orbifolds of these theories and using self consistency CFT techniques that relate the orbifold with the circle CFT at different multiples of the self-dual radius. In this section, we also introduce some more details of Matrix Quantum Mechanics in \ref{MQM} and set up our conventions. Finally in section~\ref{MMorb} we make use of the $D0$ brane picture to determine the boundary conditions of the partition function of the dual MQM, in particular we obtain the boundary conditions for the matrix $M$ and the gauge field $A$ consistent with the orbifold projection. Interestingly, we find different representations of the projection classified by an integer\footnote{This possibility was observed earlier in the unpublished work \cite{GL}.} $0 \leq n \leq N/2$. Different representations are found to be related via the action of a certain kind of ``loop-operator'' at the end-points in~\ref{loop}.
In section~\ref{canonical}, we also compute the canonical (finite N) partition function by representing it as a path integral over the eigenvalues of $M$. In addition we find that this partition function admits a natural continuation into Lorentzian signature, hence provides a possible connection to the cosmological toy universe. In particular it has a nice structure from which the initial and final wavefunctions and the transition amplitude of the toy cosmological space-time can be read off. These wavefunctions are expressed in terms of determinants of eigenvalues of $M$ at $t = 0$ and $t = T$. We further argue that the regular $n=N/2$ representation is the one expected to be dual to the orbifold in section~\ref{deconstruction}. Moreover in section~\ref{angle} we provide a dual description in terms of an angular integral with the angles corresponding to the zero modes of the gauge field $A$. \\
Section \ref{grand} is devoted to the computation of the MQM grand partition function for the ``regular" $n=N/2$  and $n=0$ representations. The grand canonical partition function is helpful in taking the double scaling limit~\cite{Boulatov:1991xz}, hence connecting the MQM partition function to the genus expansion of the dual string theory.  
This section contains one of the main findings in our paper: here we show that the calculation of the grand canonical partition reduces to the computation of the spectrum of an integral kernel which we express in various useful forms. The equations that determine the spectrum of this kernel can be expressed as integral equations. By deforming the contour of integration in these integral equations, we identify contributions to the untwisted and twisted sectors in the free energy of the orbifolded 2D NCST. \\  
It proves hard to evaluate and express these contributions in terms of the dual string theory quantities in the double scaling limit. In section~\ref{trkernel1} we perform a partial matching of the various expressions for the kernel by computing its trace, from which we can read-off the one-particle density of states that we express as a sum of the usual harmonic oscillator density of states  including a twisted state contribution. \\
Finally in section~\ref{lsegment} and in appendix~\ref{apapprox} we attempt to the twisted states at the orbifold end-points by performing a large radius expansion of the canonical partition function. We manage to do this precisely for the $n=0$ representation and the ``regular'' $n=N/2$ representation. We discover then that we can perform an exact matching with the torus contribution to twisted states computed in Liouville. The main finding of our paper is that the twisted state contribution in the scaling limit involves a Fredholm determinant of the \emph{sine-kernel} which expresses the probability that all the energy eigenvalues taken from a random Hermitian Hamiltonian lie outside the interval $[-\mu, 0]$ and thus form the fermi sea. This is also called the level spacing distribution $E_2(0, \mu)$ in the random matrix parlance~\cite{Mehta}. The initial and final wavefunctions take the form of ``square-roots'' of this distribution.\\
Throughout the text, we discuss similarities and differences with established results in the literature such as the circle and the 2D black hole~\cite{Gross:1990ub,Kazakov:2000pm}. We also discuss the possibility of realising the grand canonical partition function as a $\tau$ function of an integrable hierarchy with a Pfaffian structure.
Finally, in section~\ref{conclusions} we discuss our results and provide a look ahead. Several appendices contain the details of our calculations.

\section{The Setup}
\lab{setup}
\subsection{$c=1$ Liouville Theory on $S^{1}/{\mathbb Z}_{2}$}
\label{string}
One  computes the orbifold partition function at the torus level in string theory as folllows. Let us call the bosonic matter field\footnote{This field corresponds to the Euclidean time $\tau$ in the previous section.} $X$ restricted to the line segment $-\pi R\leq X < \pi R$ and obeying the following identifications under translation and reflection
\begin{equation}
X \approx  X+2\pi R\,\,\, and\,\,\, X \approx -X\, . 
\end{equation}
 The modular partition function of the theory is (see for example~\cite{Ginsparg:1988ui}) 
\begin{equation}\lab{orb}
Z_{orb}\left(R,z\right)=\frac{1}{2} \left\lbrace Z_{circle}\left(R,z\right)+\frac{\vert{\theta}_{2}\left(z\right){\theta}_{3}\left(z\right)\vert}{\vert{\eta}\left(z\right)\vert^2}+\frac{\vert{\theta}_{2}\left(z\right){\theta}_{4}\left(z\right)\vert}{\vert{\eta}\left(z\right)\vert^2}+\frac{\vert{\theta}_{3}\left(z\right){\theta}_{4}\left(z\right)\vert}{\vert{\eta}\left(z\right)\vert^2}\right\rbrace,
\end{equation}
where $Z_{circle}\left(R,z\right)$ is the modular partition function for the circle, $\eta\le(z\ri)$ is the Dedekind $\eta$ function, $\theta$'s are the elliptic functions, $R$ is the radius of the circle and $z$ is the modulus of the torus. The fist term in (\ref{orb}) gives the contribution from the untwisted states and equals half the partition function of the circle. The contribution from the twisted states is given by the $R$ independent part. To obtain the full torus partition function on the orbifold one should couple the ghost and the Liouville modes to (\ref{orb}) and integrate over the moduli $z$
\begin{equation}
\lab{liouv}
\mathcal{Z}_{orb}\le(R\ri)=- {V_\phi} \int_{\mathcal{F}} d^{2}z\left(\frac{\vert{{\eta}\le(z\ri)}{\vert}^{4}}{2 {z}_{2}}\right)\le(2\pi \sqrt{z _{2}}\vert \eta\le(z\ri){\vert}^{2}\ri)^{-1} Z_{orb}\le(R,z\ri)\, ,
\end{equation}
where the integral is over the fundamental domain $\mathcal{F}$, the first term in the integrand is the contribution from the ghost sector and the second the contribution from the Liouville modes. $V_\phi$ is the contribution from the Liouville zero mode, shown to be proportional to the renormalised volume in the Liouville direction $\log \mu_0$ with $\mu_0$ the renormalised string coupling~\cite{Bershadsky:1990xb}. Upon performing the integral over $z$ one finds the following answer
\begin{equation}
\mathcal{Z}_{orb}\le(R\ri)=\frac{1}{2}\mathcal{Z}_{circle}\le(R\ri)+c,
\end{equation}
where $c$ is independent of $R$ and $\mathcal{Z}_{circle}\le(R\ri)$ is the partition function of the circle coupled to the Liouville mode computed by the worldsheet methods in \cite{Bershadsky:1990xb}
\begin{equation}\lab{cir}
\mathcal{Z}_{circle}\le(R\ri)=-\frac{1}{24}\le(R+\frac{1}{R}\ri)\ln\le({\mu}_{0}\ri)\, .
\end{equation}
To determine the constant $c$ for the orbifold partition function one may use the relation between the circle and the orbifold at the self-dual radius \cite{Ginsparg:1987eb}:
\begin{equation}\lab{self}
Z_{orb}\le(R=1,z\ri)=Z_{circle}\le(R=2,z\ri)\, .
\end{equation}
Then substituting in~(\ref{self}) to (\ref{liouv}) and combining them with (\ref{cir}), one finds the final result:
\begin{equation}\label{orblfinal}
\mathcal{Z}_{orb}\le(R\ri)=-\frac{1}{48}\le(R+\frac{1}{R}\ri)\ln\le({\mu}_{0}\ri)-\frac{1}{16}\ln\le({\mu}_{0}\ri)\, .
\end{equation}

\subsection{Fermionic orbifold theories}

The classification of $\hat{c} = 1$ CFT's has been performed in \cite{Dixon:1988ac}\cite{DiFrancesco:1988xz}. According to this classification, the continuous lines of theories include two lines of ``circular" theories and various orbifolds of these theories. The ``circular" theories consist of the circle CFT and a super-affine CFT. The coupling of these ``circular" theories to super-Liouville is discussed in \cite{Douglas:2003up}. We summarize their results:
\begin{itemize}
\item \textbf{Circle CFT€™s}: The usual fermionic circle theory (compact X + Ising) gives rise to
two theories when coupled to super-Liouville: $0A$ and $0B$ depending on the GSO
projection. Their partition functions are:
\bea
\mathcal{Z}_{cirA}(R)=-\frac{1}{12\sqrt{2}}\ln\mu_0\le(2 R+\frac{1}{R}\ri),\qquad \mathcal{Z}_{cirB}(R)=-\frac{1}{12\sqrt{2}}\ln\mu_0\le( R+\frac{2}{R}\ri)\,.\nn\\
\eea
These theories are interchanged under the T-duality: $R\rightarrow 1/R$. At the special radius, $R = 1$ there is enhanced $SU(2)\times SU(2)$ symmetry.

\item \textbf{Super-Affine CFTs}: The usual super-affine theory is obtained by modding out the usual fermionic circle
theory by the following $Z_2$:
\be 
(-1)^{F_{s}}e^{2πip\delta},
\ee
where $(-1)^{F_{s}}$ is defined as +1 on the antiperiodic fermions and −1 on the periodic ones. $e^{2πip\delta}$ is a shift operator that shifts by a unit vector on the self-dual lattice.
When Coupled to super-Liouville one again obtains two theories: Super-Affine A and
Super-Affine B theories with the following partition functions:
\bea
\mathcal{Z}_{saA}(R)=-\frac{1}{12}\ln\mu_0\le( \frac{R}{\sqrt{2}}+\frac{\sqrt{2}}{R}\ri),\qquad \mathcal{Z}_{saB}(R)=-\frac{1}{24}\ln\mu_0\le( \frac{R}{\sqrt{2}}+\frac{\sqrt{2}}{R}\ri)\,. \nn \\
\eea

These theories are both self-dual under $R\rightarrow 2/R$. At the self-dual radius $R = \sqrt{2}$,
there is an enhanced $SO(3)^2$ symmetry.
\end{itemize}
Apart from the type 0 theories, there are other ``circular" $\hat{c} = 1$ theories with type I GSO projections. These have been classified in \cite{Seiberg:2005bx}. In addition to the ``circular" $\hat{c} = 1$
theories, there are three families of orbifold CFT's \cite{Dixon:1988ac}\cite{DiFrancesco:1988xz}.
\begin{itemize}
\item \textbf{Orbifold I}: The first class of orbifolds is obtained by modding out circular theories by:
\bea
R:\,\,\,\, X\rightarrow - X,\qquad \Psi \rightarrow - \Psi\,.
\eea
Both the left and right handed fermions on the world-sheet are transformed in order
to preserve world-sheet supersymmetry. $R$ as defined above is a symmetry of only
the $0B$ theory since in the $0A$ theory states in the Ramond sector have odd fermion
number. Therefore one obtains only one orbifold CFT by twisting the $0B$ theory by
$R$. 
The partition function is obtained by noting the following two relations \cite{Dixon:1988ac} which
continue to hold after coupling to super-Liouville:
\be
\mathcal{Z}_{orbI}(R)=\frac{1}{2}\mathcal{Z}_{cirB}(R)+const,
\ee
and
\be 
\mathcal{Z}_{orbB}(1)=\mathcal{Z}_{cirB}(2)
,.
\ee
The result is:
\be
\mathcal{Z}_{orbI}(R)=\frac{1}{2}\mathcal{Z}_{cirB}(R)-\frac{1}{8 \sqrt{2}}\ln\mu_0\,.
\ee

We also find two other continuous families of orbifold theories, discuss them and present their torus level partition functions in appendix~\ref{susyorb}.

\subsection{Matrix Quantum Mechanics}
\lab{MQM}
We now provide a very short review of Matrix Quantum Mechanics (MQM). For more details the reader can consult existing reviews in the literature, for example~\cite{Kazakov:1990ue,Klebanov:1991qa,Ginsparg:1993is}. Gauged MQM is a $0+1$ dimensional quantum mechanical theory of $N\times N$ Hermitian matrices denoted by $M(t)$ and a non dynamical gauge field $A(t)$. The gauge field acts as a Lagrange multiplier and projects onto the singlet representation of the $SU(N)$ gauge group. 
The path integral is defined as (we work in units where $\alpha'=1$):
\begin{equation}
\lab{inout}
\langle out | in \rangle = \int \mathcal{D} M(t) \mathcal{D} A(t) \exp\left[i N\int_{t_{in}}^{t_f} dt \mathrm{Tr} \left(\frac{1}{2}\left(D_t M \right)^2+\frac{1}{2}M^2 -\frac{\kappa}{3! \sqrt{N}}M^3\right)\right]\, , 
\end{equation}
where $D_t = \6_t + [A,M]$. 
This model has an $SU(N)$ gauge symmetry. One can diagonalise $M$ by a unitary transformation $M(t)= U(t) \Lambda(t) U^{\dagger}(t)$ where $\Lambda(t)$ is diagonal and $U(t)$ unitary.
One then picks up a Jacobian from the path integral measure for every $t$
\be \mathcal{D} M=\mathcal{D} U_{Haar} \prod_{i=1}^{N} d \lambda_i \Delta^2(\Lambda),  \ \ \
  \Delta(\Lambda)=\prod_{i <j} (\lambda_i -\lambda_j). 
\ee
This Vandermonde determinant is responsible for many interesting properties of matrix models in general and for MQM it leads to a natural description in terms of fermionic wave-functions. In particular, after projecting to the singlet sector, the Hamiltonian is found to act on the fermionic wavefunctions $\tilde{\Psi}= \Delta(\lambda) \Psi(\lambda)$ as
\be
\left(-\frac{1}{2}\frac{d^2}{d\lambda^2_i} -\half \lambda_i^2 + \frac{\sqrt{\hbar}}{3!} \lambda_i^3\right)\tilde{\Psi}(\lambda) =\hbar^{-1} E\tilde{\Psi}(\lambda), \  \quad \hbar^{-1}=\frac{N}{\kappa^2}\, ,
\ee
and describes N non interacting fermions in the cubic potential $V(\lambda)$.

To connect this model with 2D string theory one needs to send $N \rightarrow \infty$ and tune the cubic potential to a critical value $\kappa \rightarrow \kappa_c$, just before the system becomes unstable\footnote{This cubic potential is always non-perturbatively unstable, but the supersymmetric version of the model ($0B$) has a quartic stable potential and is thus non-perturbatively well defined.}. The double scaling limit is most easily performed by introducing a chemical potential $\mu$ to fill up the fermi-sea. Schematically this goes as follows: One sends $\mu, \hbar \rightarrow 0$, while keeping $\frac{\mu}{\hbar} = g_{st}^{-1}$ fixed. A careful treatment will be provided in sections~\ref{lsegment} and \ref{lbn0}. Tuning the system near the critical point is responsible for producing smooth surfaces out of the matrices~\cite{Kazakov:1990ue}. In this limit only the local maximum of the potential becomes relevant and the model thus becomes solvable, described in terms of $N$ free fermions in an inverse harmonic oscillator potential. Let us also mention that in the more modern target space approach, MQM is considered as the zero dimensional field theory living on the world volume of $N$ unstable ZZ D0- branes and the matrix field $M\le(t\ri)$ is interpreted as the open string tachyon field~\cite{McGreevy:2003kb}.

\subsection{Orbifolding in the Matrix Model Picture}\label{MMorb}

We now consider implementation of the orbifolding procedure in MQM\footnote{This procedure has been worked out in \cite{GL}.}, that corresponds to the circle orbifolding we discussed in the string theory picture above. The orbifolding procedure is very similar to the one presented in~\cite{Ramgoolam:1998vc}. We start from the Euclidean Partition function on $S^1$ (the radius is defined via $\beta_c = 2 \beta = 2 \pi R$) with the following action:
\begin{equation}\lab{D0}
S=\int_{-\beta}^{\beta}d\tau\, \tr\le(\frac{1}{2}\le(D_{\tau}M\ri)^{2} + \omega^2 M^2 \ri),
\end{equation}
where $\tau$ is the Euclidean time variable, $\beta = \pi R$, and the covariant derivative with respect to the gauge group is $D_{\tau}M={\partial}_{\tau}M-i\left[A,M\right]\,$. Here, anticipating the large N limit in (\ref{inout}) we have dropped the interaction term in (\ref{inout}) and we have allowed for a more general mass term $\omega$. For real values of $\omega$ this action corresponds to the normal harmonic oscillator potential---the inverted one can be obtained upon the analytic continuation $\omega \rightarrow i \omega$. \\
The action (\ref{D0}) is invariant under the $SU(N)$ gauge transformations:
\begin{eqnarray}\label{gauge}
M\le(\tau\ri)\rightarrow U\le(\tau\ri)M\le(\tau\ri)U^{\dagger}\le(\tau\ri),\qquad A\le(\tau\ri)\rightarrow  U\le(\tau\ri)A\le(\tau\ri)U^{\dagger}\le(\tau\ri)+i  U\le(\tau\ri){\partial}_{\tau}U^{\dagger}\le(\tau\ri).\nn\\
\end{eqnarray}
The theory also has a $\mathbb{Z}_{2}$ symmetry corresponding to 
\begin{eqnarray}
\tau\rightarrow -\tau,\qquad M\le(\tau\ri)\rightarrow M\le(-\tau\ri),\qquad A\le(\tau\ri)\rightarrow - A\le(-\tau\ri).
\end{eqnarray}
One can gauge this symmetry by projecting to the invariant states. We will use a more general gauging, by combining the reflection symmetry with a $\mathbb{Z}_2$  subgroup of the $SU(N)$ gauge group (see also~\cite{Ramgoolam:1998vc}). We define (up to a change of basis- note also that $\Omega$ is defined up to a minus sign)
\begin{equation}\label{Omega}
\Omega= \left(
\begin{array}{cc}
-1_{n \times n} & 0 \\
0 & 1_{(N-n) \times (N-n)}
\end{array} \right)*\,,
\end{equation}	
with $*f(\tau)=f(-\t)* , \ \ *\partial_\t= -\partial_\t * ,\ \ 0\leq n \leq \frac{N}{2}$,
and then require:
\begin{equation}\lab{Orb}
\Omega A(\t) \Omega^{-1} = -A(\t) + 2 n_i / \beta \delta_{i j}  ,\ \ \ \Omega M(\t) \Omega^{-1} = M(\t).
\end{equation}
with $n_i \in \mathbb{Z}$ which is allowed since the eigenvalues of $A$ are periodic variables with period $\beta$. This term turns out to be unimportant since it can be gauged-away.
This procedure naturally splits the matrices into (even/odd) blocks that need to satisfy different boundary conditions. We get
\begin{equation}\lab{block}
M(\t)= \begin{pmatrix} M_1(\t) & \Phi(\t) \\
\Phi^{\dagger}(\t) & M_2(\t) 
\end{pmatrix}, \quad 
A(\t)= \begin{pmatrix} A_1(\t) & B(\t) \\
B^{\dagger}(\t) & A_2(\t)\,. 
\end{pmatrix}
\end{equation}
One immediately sees that the $ n\times n$ $M_1 , A_1$ and the $(N- n) \times (N-n)$  $M_2 , A_2 $ matrices should be Hermitian  while the $n \times (N-n)$ $\Phi , B$ are complex. In addition, consistency with \ref{Omega}, \ref{Orb} requires that $M_1, M_2 , B$ are even while $A_1, A_2, \Phi$ are odd functions of $\t$. From the gauge transformations (\ref{gauge}) the ones that are consistent with the action of $\Omega$ are
\bea
U(\t) = \begin{pmatrix} V_1(\t) & W_1(\t) \cr
W_2(\t) & V_2(\t) 
\end{pmatrix}, 
\eea
with $V_1, V_2$ even and $W_1, W_2$ odd. \\
After orbifolding the fundamental domain is $0\leq \t \leq \beta$. In the bulk of the domain the theory is as before. The changes come from demanding different boundary conditions for the fields in~(\ref{block}) imposed at the fixed points $0, \beta$ due to their symmetry. In particular we need to demand
\be
A_1(0)=A_2(0)=\Phi(0)=0=A_1(\beta)=A_2(\beta)=\Phi(\beta)
\ee
The $SU(N)$ gauge group gets broken to $SU(n) \times SU(N-n)$ at the boundaries, and as we will see the initial and final wavefunctions contain two separate sets of $n$ and $N-n$ fermions.
This breaking also means that the zero-modes of $A$ come solely from the off-diagonal elements $B$.

The two most special cases are\footnote{From now on we assume even $N$.} $n=0$ and $n=N/2$. The first is the simplest case where there exist no zero modes of the gauge field, while the second describes the so-called ``regular" representation of the orbifold which is expected to give the Matrix Model dual to the orbifold Liouville theory. Further reasoning for why $n=N/2$ is expected to be the correct representation, based on ideas related to deconstruction, is provided in section~\ref{deconstruction}.

Representations with different $n$ correspond to adding fractional D-instantons at the fixed points. This becomes clear in the T-dual picture. In particular, upon T-dualizing the Euclidean circle to a radius $1/R$ and then orbifolding, the original $D_0$ branes become D-instantons whose position on the dual circle is governed by the zero mode of the gauge-field. This means that for $n=0$, $A(\tau)=n_i \frac{\pi}{R} \delta_{i j}, \, n_i \in \mathbb{Z}$ and all the instantons are stuck at the fixed points $0, \pi/R$. For the generic $n$ representation, one has $n$-zero modes with arbitrary angle in the T-dual circle and thus the configuration contains $n$-physical instantons at angles $\theta_i$ together with $N-2n$ stuck at the fixed points. This makes clear that the regular $n=N/2$ representation has only physical instantons in the T-dual picture. More discussion about how to connect different representations will follow in section~\ref{loop}.

\section{The Canonical Partition Function}
\label{canonical}
The partition function for a generic n representation of the orbifold is then obtained by integrating over the non-vanishing components of the matrices $M$ and $A$ in (\ref{block}) at the initial and final points---that are the even components $M_1$, $M_2$ and $B$ at $\tau=0$ and $\tau=\beta$, and performing the path integral of the full matrices between these points. Thus, for a generic n-representation we have, 
\be\label{norb}
\mathcal{Z}= \int \mathcal{D} B(0) \mathcal{D}B(\beta) \mathcal{D} M_{1,2}(0) \mathcal{D}M_{1,2}(0) \int_{A(0)}^{A(\beta)} \mathcal{D}A(\t) \int_{M(0)}^{M(\beta)} \mathcal{D}M(\t) e^{-S}\,.
\ee
The next step is to reduce this matrix integral to an integral over eigenvalues. One can show that
\be\label{gaugedprop}
\int_{M(0)}^{M'(\beta)} \mathcal{D}A \mathcal{D}M e^{-\int_0^\beta d\tau \mathrm{Tr} \half (D_\tau M)^2 + \half \omega M^2} = \int_{U(N)} \mathcal{D} U \langle U M' U^\dagger, \beta | M, 0 \rangle\,.
\ee
In our case the propagator is the (Euclidean) propagator for a matrix harmonic oscillator given by 
\begin{equation}\label{matrixprop}
\resizebox{.9\hsize}{!}{$\langle M', \beta | M, 0 \rangle = \left(\frac{\omega}{2\pi \sinh \omega \beta} \right)^{N^2/2} \exp\left(-\frac{\omega}{2 \sinh \omega \beta}\left[\left(\mathrm{Tr} M^2 + \mathrm{Tr} M'^2 \right) \cosh \omega \beta - 2 \mathrm{Tr} M M'   \right]\right)$}
\end{equation}
These two equations can be combined beautifully using the Harish-Chandra-Itzykson-Zuber integral
\be\label{HCIZ}
\int_{U(N)} DU \exp{\left( g \mathrm{Tr} M U M' U^\dagger \right) } = \prod_{p=1}^{N-1}( p!) \ g^{-\half N (N-1)} \frac{\det e^{g \lambda_i \lambda_j'}}{\Delta(\lambda) \Delta(\lambda')} 
\ee
that will allow us to reduce the integral to eigenvalues. This is possible since the only term that couples different matrices in the propagator is precisely of the form that can be reduced to eigenvalues via the HCIZ formula.

Before moving on, let us note the following two options: we can either first diagonalise $M$ and then integrate over $U$ using the HCIZ formula or first diagonalise $U$ and then integrate over $M$. In the orbifold case one also needs to take care about the orbifold projection which is implemented through the block structure of the matrices. In the next section, we will follow the first procedure and compare the results for the circle and orbifold. In section~\ref{angle} we will follow the second and in section~\ref{trkernel1} we will perform a matching between the two methods.

\subsection{Partition function in terms of eigenvalues}

To set up our notation, we define $K^E(\lambda_i, \lambda'_j; \beta)= \langle \lambda_j', \beta | \lambda_i, 0 \rangle$ the Euclidean oscillator propagator as follows
\bea
\label{upmehler}
K^E(\lambda_i, \lambda'_{j} ; \beta)&= \left(\frac{\omega}{2\pi \sinh \omega \beta} \right)^{\half} \exp\left(-\frac{\omega}{2 \sinh \omega \beta}\left[\left( \lambda_i^2 +  {\lambda'}_{j}^{2} \right) \cosh \omega \beta - 2  \lambda_i \lambda'_{j}  \right] \right) \nn \\
 &= \sum_{n=0}^\infty \psi_n (\lambda_i) \psi_n (\lambda'_{j}) q^{n+\half},
\eea
where the second spectral representation is also known as Mehler's formula. In this representation $q=e^{- \omega \beta}$ and $\psi_n (\lambda_i)$ are the Hermite functions. Note that upon analytic continuation $\omega \rightarrow i \omega$ the Hermite functions turn into parabolic cylinder functions $D_\nu (z)$ defined for complex $\nu, z$ see Appendix~\ref{parabolic}. One can also resolve the inverted oscillator propagator in terms of parabolic cylinder functions from the start~\cite{Moore:1991sf}, the relevant formula is presented in Appendix~\ref{Mehler}. As we discuss below the possibility of analytically continuing the propagator in the parameters $\omega , \beta$ is the reason we expect to obtain the Lorentzian transition amplitude in this 2D toy universe directly from  the Euclidean description.

\subsubsection{The circle}
We first review the case of circle~\cite{Gross:1990ub,Boulatov:1991xz}. For this partition function on  $S^1$ we just have to demand periodic boundary conditions ($M'(\beta_c)=M(0)$, $\beta_c = 2 \beta$)
\be\label{circle1}
\mathcal{Z}_N=\int \mathcal{D} M(0) \mathcal{D}U_1 \langle U_1 M(0) U_1^\dagger | M(0) \rangle =\frac{1}{N!} \int \prod_{i=1}^N d \lambda_i \det K^E(\lambda_i, \lambda_j),
\ee
where we diagonalised $M(0)= U_2 \Lambda U_2^\dagger$ and integrated over the matrix $U= U_1 U_2^\dagger$. The $\Delta^2(\Lambda)$ in the numerator from the measure of M, canceled the similar term produced by the HCIZ formula. The term $\prod_{p=0}^{N-1} p!$ in the HCIZ formula got canceled by the second integration over the gauge group which is $\int \mathcal{D} U_{Haar}= \pi^{N(N-1)/2}/\prod_{p=0}^{N} p!$. In the end the $1/N!$ term is due to the left-over permutation (Weyl) symmetry between the eigenvalues. For more details on factors of N for this and more general cases see~\cite{Eynard:2015aea}.\\
The result is the partition function of N free fermions in the harmonic oscillator potential: 
\be
\label{cpfcircle}
\mathcal{Z}_N = \frac{q^{\frac{N^2}{2}}}{\prod_{k=1}^N (1-q^k)},
\ee
with $q=e^{- \omega \beta_c}$. One can also expand this result for large $\beta_c$ and recover the zero temperature free energy
\be
\mathcal{F} = \beta_c\, \omega\,  \frac{N^2}{2} = \beta_c E_0 + O(e^{-\omega \beta_c}),
\ee
where $E_0= \sum_{k=0}^{N-1} \omega (k +\half)$ the vacuum energy of the system of N fermions.
In constrast, in the orbifold case at least for the torus contribution we expect a subleading $\beta$ independent term due to the presence of twisted states since these are localised at the end-points.

\subsubsection{The orbifold partition function for generic n}
The orbifold partition function for generic n after we integrate over the propagation becomes
\be
\mathcal{Z}_{n, N-n}= \int \mathcal{D}M \mathcal{D}M' \mathcal{D}U \langle U M' U^\dagger, \beta | M, 0 \rangle,
\ee
with \be M=\begin{pmatrix}
M_1^{(n \times n)} & 0 \\ 0 & M_2^{(N-n) \times (N-n)}
\end{pmatrix}, \ \quad \mathcal{D}M=\mathcal{D}M_1 \mathcal{D}M_2 \ee 
and similarly for $M'$. 
We now use the HCIZ formula to evaluate the integral over the unitary matrix $U$. If we define the eigenvalues of $M_{1,2}$ as $x_i ,y_i$ respectively, $\prod_{i=1}^n d x_i /n! \equiv d^n x$ and similarly for $y$, the result is found to be
\be
\label{eigenvaluerepn}
\mathcal{Z}_{n, N-n}= C_{N, n} \int d^n x d^{N-n} y d^n x' d^{N-n} y'  \frac{\Delta_{n}(x) \Delta_{N-n}(y)}{\Delta_{n,N-n}(x, y)} \det K(\bar{x}_i ; \bar{x}_j') \frac{\Delta_n(x') \Delta_{N-n}(y')}{\Delta_{n,N-n}(x', y')},
\ee
with
\be
\Delta_{n,N-n}(x, y) = \prod_{i =1 }^{n} \prod_{j =1 }^{N-n} (x_i - y_j)\,.
\ee
First of all we make the following crucial observation: the form of this Euclidean partition functions in~(\ref{norb}) and~(\ref{eigenvaluerepn}) are appropriate for analytic continuation into the Lorentzian time. The analytic continuation is obtained simply by changing $\beta = i T$ in the propagator. Therefore, after the analytic continuation we can simply interpret these Lorentzian partition functions as "transition amplitudes" from an initial state of the universe at $t = 0$ to a final state at $t = T$ 
\be\lab{decomp}
\langle \psi_f , T | \psi_i 0 \rangle = \int D\bar{x} D\bar{x}^{'} \psi^*_f (\bar{x}^{'}) \det K^L (\bar{x}, \bar{x}^{'}; T) \psi_i (\bar{x})\, ,
\ee
where we introduced the compact notation $\bar{x}=(x,y)$. Here the initial and final wave-functions in this toy universe are of the form 
\be\lab{inifin}
\psi_i(\bar{x}) = \psi_f(\bar{x}) = \frac{\Delta_{n}(x) \Delta_{N-n}(y)}{\Delta_{n,N-n}(x, y)}\, .
\ee  
One can rewrite these wavefunctions in the form $\prod_{i, j} (\lambda_i - \lambda_j)^{q_i q_j}$ in terms of fermions having positive ($q_i= +1 \, , 1\leq i \leq n $) and negative  ($q_i= -1 \, , n+1 \leq i \leq N $) charge (or spin), with same charge fermions ``feeling" repulsion and opposite ones attraction. They represent a Coulomb-gas in one dimension. From this point of view, the representation $n=N/2$ is the only one satisfying charge neutrality. Wavefunctions of this form first arised in studies of Quiver Matrix Models~\cite{Kostov:1995xw,Kharchev:1992iv,Kazakov:1998ji}\footnote{In some of these studies the divergence coming from the denominator is avoided, since it has the form $x_i + y_j$ with the variables restricted to be positive. We will regulate this divergence taking the principal value in section~\ref{regulargrand}.}. Then they reappeared in connection to the description of effective IR superpotentials of $\mathcal{N}=1$ gauge theories and in studies of supermatrix models (see~\cite{Dijkgraaf:2002dh,Dijkgraaf:2002vw,Dijkgraaf:2003xk} and references within). If one replaces rational with hyperbolic functions, a similar ratio can also be found in studies of superconformal Chern-Simons theories of Affine $\hat D$-type, at the quiver end-nodes~\cite{Moriyama:2015jsa}. Finally there is recent interest in these wavefunctions \cite{Vafa:2014iua} in the context of non-Unitary holography.

\subsubsection{Changing representations via Loop operators}\label{loop}
One can connect different n representations by inserting operators in the end-points of the path integral of the form $\prod_j (x - y_j)^2$ or $\prod_j 1/(x- y_j)^2$ to lower/raise the value of n.
This form of operators is known as loop operators. We first define the loop operator that creates macroscopic holes/boundaries on the worldsheet (this means the string gets attached to a D-brane, the so called FZZT brane) in matrix model language~\cite{Moore:1991ir,Ginsparg:1993is,Martinec:2004td,Kutasov:2004fg,Maldacena:2004sn}:
\begin{equation}
W(x)=\frac{1}{N} \tr \log (x- M)\,.
\end{equation}
The function that creates a coherent state of them is:
\begin{equation}
e^{N W(x)} = \det(x-M)=\prod_{j}^{N} (x - \lambda_j)\,.
\end{equation}
In these equations x can be thought of as a chemical potential $\mu_B$ (or a boundary cosmological constant). For $c<1$ theories these operators have been thoroughly studied from the matrix model point of view in~\cite{Kutasov:2004fg,Maldacena:2004sn}. The relevant branes are the FZZT branes which extend along the Liouville direction. In our case let us take as an example the operator that transforms the generic $n$ to the $n=0$ representation
\be
\prod_i^n \prod_j^{N-n} (x_i - y_j)^2 = \det\left(M_1 \otimes \unit_{N-n \times N-n} - \unit_{n \times n} \otimes M_2 \right)^2,
\ee
where the determinant is in the tensor product space. Similarly one can transform the generic $n$ representation to the $n=N/2$ by an inverse determinant of the same form. From this expression, it is easy to see that the eigenvalues of $M_1$ act as chemical potentials for the eigenvalues of $M_2$ and vice versa and have to be integrated over (they do not represent external parameters as in the familiar computations that involve FZZT branes). The open strings are the ones stretched between the two sets of $n$ and $N-n$ $D0$ branes which can be thought to separate at the end-points due to the breaking $SU(N) \rightarrow SU(n) \times SU(N-n)$. 
Using a Miwa-style representation~\cite{Morozov:1994hh}, the authors of~\cite{Mukherjee:2005aq} found that in the case of the Normal matrix model, these determinants/inverse determinants  decrease/increase the closed string tachyon coupling thus deforming the closed string background. Therefore there exist two complementary ways to understand these operators (open/closed duality). In our case let us note that similarly we can write
\be
\resizebox{.93\hsize}{!}{$\det\left(M_1 \otimes \unit_{N-n \times N-n} - \unit_{n \times n} \otimes M_2 \right)^2= e^{\left [ (N-n) \tr \log M_1 + n \tr \log M_2  - \sum_{k=1}^\infty t_k M_2^k  - \sum_{k=1}^\infty \bar{t}_k M_1^k \right]}$}
\ee
where we chose to expand each determinant factor in a different way and $t_k = \tr( M_1^{-k} ) /k$, $\bar{t}_k = \tr( M_2^{-k} ) /k$. These are the closed string tachyon couplings in Miwa variables. The logarithmic terms in the exponent appear in versions of the Penner model, for more details one can consult~\cite{Mukhi:2003sz}. This description makes clear that there is a backreaction effect where $M_1$ deforms the closed string background of $M_2$ and vice versa. \\
One might furthermore try to use grassmannian/fermionic variables to exponentiate these factors~\cite{Kutasov:2004fg,Maldacena:2004sn}. In particular we get
\be
\label{fermionsloop}
\det\left(M_1 \otimes \unit_{N-n \times N-n} - \unit_{n \times n} \otimes M_2 \right)= \int d \chi^\dagger d \chi e^{\chi^\dagger \left(M_1 \otimes \unit_{N-n \times N-n} - \unit_{n \times n} \otimes M_2 \right) \chi},
\ee
with $\chi_{\alpha j}, \chi_{\alpha j}^\dagger$ fermions transforming in the bifundamental representation of $SU(n) \times SU(N-n)$\footnote{Integrating-in fundamental fermions had been already used in the context of $c=1$ open string theory in~\cite{Minahan:1992bz}.} that exist only at the orbifold endpoints. One could also endow these fermions with a kinetic term (dynamic-loops on the worldsheet) as in~\cite{Minahan:1992bz}, that would correspond to the T-dual picture (Neumann conditions in Euclidean time for open strings). This construction also indicates that determinants correspond to fermionic open strings streched between the branes, while inverse determinants to bosonic open strings~\cite{Kutasov:2004fg,Maldacena:2004sn,Mukherjee:2005aq}. It would be very interesting to study further our model from this point of view and connect it with various ideas related to FZZT branes in the existing literature and possibly understand non-perturbative effects as well.



\subsubsection{The $n=0$ case}
\lab{pfn=0}
This is the simplest case, where the zero modes of the gauge field vanish. The line segment partition function for $n=0$ has a  structure similar to the two-matrix model~\cite{Mehta:1981xt,Kazakov:1986hu}.
\be
\mathcal{Z}_{0,N}= \int \mathcal{D}M \mathcal{D}M'  \langle  M' , \beta | M, 0 \rangle = C_N \int \prod_{i=1}^N d \lambda_i d \lambda_i' \Delta(\lambda') \det_{i, j} K^E(\lambda_i; \lambda_j') \Delta(\lambda),
\ee
with $C_N$ a constant. 
One can also compute the canonical partition in this case using the methods in the appendix of~\cite{Mehta:1981xt} or by direct Gaussian integration to find 
\be
\label{canonicaln0}
\mathcal{Z}_{0,N}= \left( \frac{2 \pi}{\omega \sinh \omega \beta } \right)^{N^2/2}\,.
\ee
Defining the partition function of a single harmonic oscillator with open boundary conditions~\cite{Kleinert}
\be\label{1popen}
\mathcal{Z}_{1}^{op}= \int_{-\infty}^\infty dx  dx' \langle x | x'\rangle = \left(\frac{2 \pi}{\omega \sinh \omega \beta}\right)^\half ,
\ee
one finds that the $n=0$ partition function is just $N^2$ copies of the single particle one
\be
\mathcal{Z}_{0,N}= \left( \mathcal{Z}_{1}^{op} \right)^{N^2}\,.
\ee
For large $\beta= \beta_c/2$ one again obtains
\be
\mathcal{F}_{0,N} = \half \beta_c E_0 +  \frac{N^2}{2} \log C + O(e^{-\omega \beta_c}),
\ee
with the $\beta$ independent term depending on the normalization of the partition function. We evaluate this $\beta$-independent term from the canonical partition function in an unambiguous manner in section \ref{lsegment} and try to directly perform the double scaling limit.

\subsubsection{The $n=N/2$ Case}
\lab{pfn=N/2}

As we discussed, the regular $n=N/2$ case is special and expected to be the correct dual of the 2D string theory on the orbifold, see also the discussion in section~\ref{deconstruction}.
In order to facilitate the computation of the partition function in this case, let us first introduce the Cauchy identity
\be\label{cauchy}
\frac{\prod_{i<j}(x_i-x_j)\prod_{i<j}(y_i-y_j)}{\prod_{i,j}(x_i-y_j)}
=\det\frac{1}{x_i-y_j}\, .
\ee
Using this identity we can express the $n=N/2$ partition function as
\be\label{2x2matrixpf}
\mathcal{Z}_n = C_{n} \int d^n \bar{x} d^n \bar{x}' 
\det_{n \times n}\left( \frac{1}{x_i-y_j} \right)
\det_{2n \times 2n} \begin{pmatrix} K^{n}(x_i, x'_j) & K^{n}(x_i, y'_j) \\ K^{n}(y_i, x'_j) & K^{n}(y_i, y'_j) \end{pmatrix}
\det_{n \times n}\left( \frac{1}{x'_i-y'_j} \right)\,.
\ee
Now one can perform  the integration over $(x', y')$ using the extension of Andr\' eief's identity for matrices of different ranks presented in \cite{Matsumoto:2013nya}. The result is
\begin{align}
\mathcal{Z}_n \sim (-1)^n \int d^n x d^n x'
\det\begin{pmatrix}
K(x_i,x'_j)&(K\bullet N)(x_i,x'_j)\\
(M\bullet K)(x_i,x'_j)&(M\bullet K\bullet N)(x_i,x'_j)
\end{pmatrix},
\end{align}
where $\bullet$ stands for either the $y$ integration or the $y'$ integration and we defined $M(x_i,y)= \frac{1}{x_i-y}, \ N(y', x_j)= \frac{1}{y'-x'_j}$ to distinguish between these two cases. For example
\be
(M\bullet K\bullet N)(x_i,x'_j) = \int dy dy' M(x_i, y)K(y, y')N(y', x'_j)\,.
\ee
Thus, in this way, we manage to trade integrals over $n$ variables with integrals over single variables.\\ 
One can also perform the $x'$ integrations using a formula by de Bruijn~\cite{deBruijn:1955,Moriyama:2015jsa} to get
\begin{align}
\label{pfaffianpf}
\mathcal{Z}_n \sim(-1)^{n+\frac{1}{2}(n-1)n}\int d^n x
\pf  P,\quad P=\begin{pmatrix}P_{11}&P_{12}\\P_{21}&P_{22}\end{pmatrix}\, ,
\end{align}
where ``pf'' stands for the Pfaffian and the four $n\times n$ blocks given by
\begin{align}
P_{11}
&=-(K\circ N\bullet K+K\bullet N\circ K),
&P_{12}
&=(K\circ N\bullet K+K\bullet N\circ K)\bullet M,
\nonumber\\
P_{21}
&=-M\bullet
(K\circ N\bullet K+K\bullet N\circ K),
&P_{22}
&=M\bullet
(K\circ N\bullet K+K\bullet N\circ K)\bullet M\,.
\end{align}
with the $\circ$ standing for integration over $x'$. Let us note that the Pfaffian structure we find here is very similar to the one encountered in studies of the affine $\hat D$ superconformal Chern-Simons theory in
\cite{Moriyama:2015jsa} and we followed essentially the same steps in deriving the equation~(\ref{pfaffianpf}).

\subsubsection{Deconstruction and Quiver Matrix Models}\label{deconstruction}

There is another important reasoning on why one expects the regular $n=N/2$ representation to be the one related to the orbifold on the Liouville theory. It is based on deconstruction arguments in relation to the study of $c=1$ CFTs at multiples of the self-dual radius. \\
 In particular a survey of $c=1$ CFT's shows that the orbifold CFT is related to the critical Ashkin-Teller model \cite{Saleur:1987,Ginsparg:1987eb,Dijkgraaf:1987vp} that describes two Ising spins coupled by a four spin interaction. The theory at a multiple $m R^o_{sd}$ of the self-dual radius $R^o_{sd}$ has an affine $\hat{D}_m$ symmetry. The reasoning for this is analogous to the one for the affine $\hat{A}_{2m-1}$ symmetry of the circle theory at multiples of the circle self-dual radius $m R^c_{sd}= 2 m R^o_{sd}$ and based on studying string propagation on $SU(2)/\Gamma$ with $\Gamma \subset SU(2)$ a finite subgroup of $SU(2)$ (the binary cyclic group $\mathcal{C}_{2n}$ for the circle and the binary dihedral group $\mathcal{D}_n$ for the orbifold). 
 
\begin{figure}[t]
\vskip 10pt
\centering
\includegraphics[width=120mm]{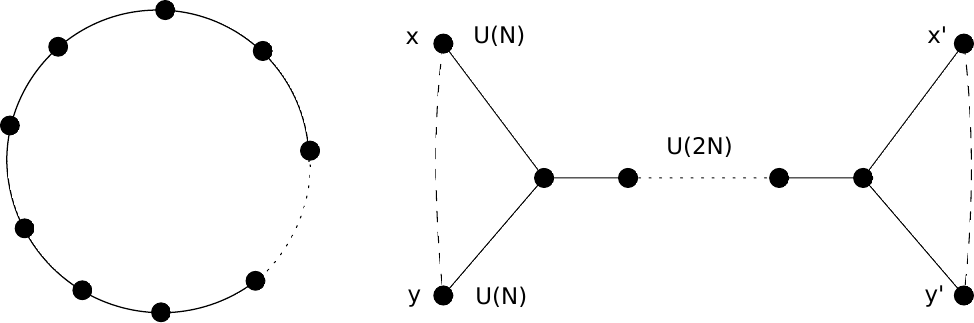}
\caption{The affine $\hat{A}$ and $\hat{D}$ quivers. For the $\hat{D}$ case the nodes in the middle correspond to $U(2N)$, while the end-nodes to $U(N)$ adjoint fields. The connections between the nodes correspond to bi-fundamental fields. In studies of superconformal-chern-simons theories~\cite{Moriyama:2015jsa}, it was found that one can write the partition function integrand as a product of determinants with the end-nodes interacting pairwise as in eq.~\ref{dquiver}.}
\label{fig:quiver}
\end{figure}

Deconstruction is a form of discretization of a continuous dimension pioneered in~\cite{ArkaniHamed:2001ca}. We will be mostly interested in the proposed description of $c=1$ string theory, where the dual matrix description for multiples of the self-dual radius on the circle is in terms of an $\hat{A}_{2m+1}$ quiver of matrices and
that for the orbifold should similarly expected to be in terms of a $\hat{D}_m$ quiver of matrices~\cite{Kostov:1992ie,Dijkgraaf:2003xk}(see fig.~\ref{fig:quiver}). Moreover, in the case of $\hat{A}$-quiver, the partition function can typically be written as the integral of a determinant~\cite{Kostov:1995xw}, a fact encountered also in the study of the $\hat{A}$ matrix-quiver of superconformal Chern-Simons theory, where the rational functions are replaced by hyperbolic functions~\cite{Marino:2011eh}. As we saw in equation~(\ref{circle1}), this is also true for the $S^1$ partition function for arbitrary radius. Similarly it is known from studies of the ABJM $\hat{D}$-type quiver matrix model, that the partition function has the structure of a Pfaffian~\cite{Moriyama:2015jsa}. In particular at the end-nodes $x, y$ of the $\hat{D}$ quiver where the gauge group breaks from $U(2N)$ to $U(N)\times U(N)$\footnote{A non-symmetric breaking should probably be understood as containing extra non-perturbative effects which is consistent with the picture of adding D-instantons at the end-points.}, one finds a factor
\be
\label{dquiver}
\frac{\prod_{i<j} \sinh\left( \frac{x_i - x_j}{2}\right) \sinh\left(\frac{y_i-y_j}{2}\right)}{\prod_{i , j}\sinh\left(\frac{x_i - y_j}{2}\right)} = \det \frac{1}{\sinh\left(\frac{x_i - y_j}{2}\right)}
\ee
which is very similar to our expression (\ref{inifin}) where the rational functions are replaced by hyperbolic ones.  In addition, for the rational case the limit of an infinite number of nodes ($\hat{A}_\infty$ -case), is expected to describe the $c=1$ infinite line as a limit of the circle at infinite radius and similarly one should expect the infinite line with twisted states at the endpoints to be obtained either taking an infinite radius limit for the orbifold, or equivalently by studying the $\hat{D}_\infty$ quiver.

In the light of this discussion, it becomes more clear why the regular representation should contain the correct description of the orbifold, as it shares the same symmetry breaking at the end-points with the $\hat{D}_m$ quiver-matrix model. Moreover it should also match with it at the corresponding multiples of the self-dual radius. On the other hand this also explains why we were able to write the integrand of the partition function as a Pfaffian. A final property that singles out the regular representation is that this is the only case where the wavefunctions at the endpoints are square integrable on the infinite line $x, y \in (-\infty,\infty)$. In particular taking the generic wavefunction
\be
\psi_{n, N}(x, y) = \frac{\Delta_{n}(x) \Delta_{N-n}(y)}{\Delta_{n,N-n}(x, y)} 
\ee
we find that it scales as 
\bea
\psi_{n,N} \sim x_i^{2n-N-1}, \quad for \quad x_i \rightarrow \infty \nn \\
\psi_{n,N} \sim y_j^{N-2n-1}, \quad for \quad y_j \rightarrow \infty
\eea
and thus can be square integrable, $\int |\psi |^2 d^n x d^{N-n} y < \infty$, only for the representation $n=N/2$\footnote{The divergences at the locii $x_i = y_j$ are regulated below by adopting the principal value prescription.}.

\subsection{Canonical partition function in terms of angles}\label{angle}

In this section we will follow the other possible method of evaluation of the canonical partition function that involves diagonalising the unitary matrix $U$ and integrating over $M$.

\subsubsection{The circle}

We start again by reviewing the circle case. In equation~(\ref{circle1}) we diagonalise $U_1$ by a unitary transformation as $U_2 U_1 U_2^\dagger= \delta_{ij} e^{i \theta_j}$ and integrate over $M(0)$ in the path integral to obtain~\cite{Boulatov:1991xz}:
\bea\label{circle2}
\mathcal{Z}_N= \resizebox{.32\hsize}{!}{$\int \mathcal{D} M(0) \mathcal{D}U_1 \langle U_1 M(0) U_1^\dagger | M(0) \rangle$} & =& \frac{1}{N!} \int_0^{2 \pi} \prod_{k=1}^N \frac{ d \theta_k}{2 \pi} |\Delta(e^{i \theta})|^2 q^{\half N^2} \prod_{i j} \frac{1}{1 - q e^{i(\theta_i - \theta_j)}}  \nn \\
& = & \frac{1}{N!} \oint \prod_{k=1}^N \frac{ d z_k}{2 \pi i} \det_{i,j} \frac{1}{q^{\half} z_i - q^{-\half} z_j},
\eea
with $z_i = e^{i \theta_i}$, $q=q_c = e^{- \omega \beta_c}$ and in the second line we used the Cauchy identity. The result of the integrations was found~\cite{Boulatov:1991xz} to agree with~(\ref{cpfcircle}).

\subsubsection{The orbifold for generic n}

We now write down the result for the generic n-representation of the orbifold \cite{GL}. The details of this calculation are presented in the appendix~\ref{anglerep}.
\be
\label{ncanonical}
\mathcal{Z}_n = \int_0^\pi \prod_k d \theta_k J_n(\theta) I_n(\theta)
\ee
with ($\mathcal{Z}_{1}^{o}$ given by eqn.~\ref{1popen}) \small
\be
 I_n= \left( \mathcal{Z}_{1}^{o} \right)^{\frac{ (N-2n)^2}{2}} \prod_i^n\left[\frac{2}{\cosh \tilde{\beta} - \cos  \theta_i} \right]^{N-2n} \prod_{i,j}^n\left[ \frac{4}{(\cosh \tilde{\beta} - \cos (\theta_i+\theta_j) (\cosh \tilde{\beta} -  \cos( \theta_i-\theta_j)} \right]^{\half} 
\ee 
\normalsize where $\tilde{\beta}=\omega \beta_c = 2 \omega \beta$  and the measure
\be
J_n(\theta)=\frac{1}{2^n n! (2 \pi)^n} \prod_{i<j}^n \sin^2 \left(\frac{\theta_i - \theta_j}{2} \right) \sin^2 \left(\frac{\theta_i +\theta_j}{2} \right) \prod_{k=1}^n \sin \theta_k \sin^{2(N-2n)}\left(\frac{\theta_k}{2} \right)\,. 
\ee
This looks quite complicated, but as we show in appendix~\ref{anglerep}, it can be expressed as the inverse quarter of determinant of the differential operator $Q$
\be \label{IQ1}
I=\left(\frac{2\pi}{\o}\right)^{\frac{1}{2}(N-2n)^2}(\det Q)^{-\frac{1}{4}}\, ,
\ee
where
\be
Q=- D_{\tau}^2+{\o}^2,\qquad D_\tau = \6_\tau + i[A, \cdot]\, ,
\ee
with a diagonal constant gauge field 
\be\lab{wilsonorb}
A=(\pi R)^{-1}\textrm{diag}(\q_{1},\q_{2},...,\q_{n}, -\q_{1},-\q_{2},...,-\q_{n},0,...,0)\, .
\ee
This form makes it clear that the angles $\q_i$ can be interpreted in the T-dual picture as the positions of the D-instantons that can move freely. The vanishing $\q_i$ for $i=2n+1, \cdots N$ correspond to the fractional D-instantons that are stuck at the fixed points. The extra power $1/2$ in (\ref{IQ1}) is due to the orbifold projection.

We also show below that the particular case of $n=N/2$ enjoys a nice Pfaffian structure. Furthermore as a consistency check, one obtains the previous expression~(\ref{canonicaln0})  for $\mathcal{Z}_{n=0}$, since in this case there is no integral to be performed and therefore one just picks up the prefactor. 
Finally, the large $\beta$ expansion for generic n can be found in Appendix~\ref{lbgeneric}.

\subsubsection{The orbifold for $n=N/2$}\label{angleorbifold}

For the special $n=N/2$ representation one finds that the canonical partition function can be written in terms of a Pfaffian, as derived in Appendix~\ref{Pfangles}:
\be
\label{anglepf}
\mathcal{Z}_n= \frac{1}{n !} \int_0^\pi \prod_{k=1}^n \frac{ d \theta_k }{2\pi i} \prod_{k=1}^n \frac{  q^\half}{\sqrt{\left(1-q z_k^2 \right)\left(1-q {z^*_k}^2\right)}} \pf \begin{pmatrix}\frac{q^{1/2} (z_i- z_j)}{1-q z_i z_j}&\frac{q^{1/2} (z_i- z^*_j)}{1-q z_i z^*_j}\\\frac{q^{1/2} (z^*_i- z_j)}{1-q z^*_i z_j}&\frac{q^{1/2} (z^*_i- z^*_j)}{1-q z^*_i z^*_j}\end{pmatrix},
\ee
with $z_i= e^{i \theta_i}$, $q=e^{- \omega \beta_c}$.\\
This expression is very interesting. We notice that the terms in the measure take values around the full circle and the square-root leads to branch cuts in the complex $z_i$ plane. One can also exponentiate the measure to obtain an equivalent expression
\be
\mathcal{Z}_n=  \frac{q^{\frac{n}{2}}}{n !} \int_C \prod_{k=1}^n \frac{ d z_k }{2 \pi z_k} \prod_{k=1}^n e^{\sum_{j=1}^\infty t_j z_k^{2j} + \sum_{j=1}^\infty t_{-j} z_k^{-2j} } \pf P, 
\ee
with $t_j = t_{-j} = q^j/2j$ and the contour $C$ is the upper-half plane semi-circle. \\ Comparing this expression with the analogous matrix model description of the 2D black-hole (see~\cite{Kazakov:2000pm}) one notices that the couplings $t_j$ act by turning on vortex perturbations or Wilson-lines ($t_k \tr U^k$) whose strength in our case is determined by the inverse temperature $\beta$\footnote{In that case only $t_{\pm 1}$ were turned on and were independent parameters of the model.}. Moreover, in our case, all these couplings are related and at large $\beta$ the most relevant ones are $t_{\pm 1}$ which vanish as $e^{- \omega \beta_c}$. This type of perturbations can be encountered also in variants of the Gross-Witten-Wadia model~\cite{Gross:1980he,Jurkiewicz:1982iz,Periwal:1990gf,Mironov:1994mv,Hoppe:1999xg} some of which are known to exhibit phase transitions. The form of these couplings thus raises the possibility of encountering a phase transition as one lowers $\beta$, but one should keep in mind that for the relevant case of the inverse harmonic oscillator,  $\omega=-i, \, \, q=e^{i \beta_c}$ corresponds to a phase and a more careful study is needed. We should also mention here that based on a generalised version of the FZZ-duality, it is conjectured that higher-windings are related to higher spin generalisations of the 2d black hole~\cite{Mukherji:1991kz}, where discrete states are liberated as well~\cite{Mukherjee:2006zv}.\\

Let us finally comment on the possibility of relating this partition function to an integrable hierarchy\footnote{We wish to thank A.Morozov, A.Yu.Orlov and J.van de Leur for discussions related to this possibility.}. If true, this would, on the one hand, indicate that the model is integrable, and on the other hand it would provide us with differential equations for the partition function in terms of its parameters, as in the case of the 2D black hole~\cite{Kazakov:2000pm}. To this end, we first note that the couplings $t_k$ act as deformation parameters in a Miwa parametrization. The next step is identifying an appropriate $2n$-free fermion correlator that gives the specific Pfaffian. This Pfaffian structure for free fermions is encountered in BKP/DKP hierarchies, for more information the reader can consult~\cite{Jimbo:1983if,vandeLeur:2015,Orlov:2016} and the references therein. The most important and final difficulty is the fact that the integration is not around the circle but from 0 to $\pi$, thus one needs free fermionic correlators with branch cuts like in the Ramond sector. Since the fermionic modes are expanded in semi-integer powers, this means that the fermions live on the double-cover of the $z$ plane, or equivalently in the background of a twisting field, that creates ramification points. In section~\ref{ellfpar} we find that the natural way to understand the double cover---that turns out to be a torus---is by using a parametrization in terms of Jacobi's elliptic functions.

\subsubsection{A non-perturbative symmetry}

It is also easy to check that the partition function for $n=N/2$ admits an exact symmetry upon rotating $\o \rightarrow - i \o$ and $\beta \rightarrow i T$ together. This is because the partition function depends only on the product $\omega \beta$\footnote{In fact it is also invariant under flipping the sign of $\omega \beta$.}. This can be seen either from eqn.~\ref{2x2matrixpf} by rescaling the matrix eigenvalues or directly from eqn.~\ref{I}. This property is also shared with the $S^1$ partition function and furthermore does not hold for any of the other $n$, that nevertheless just pick phase factors that depend on $n$.
 
This symmetry indicates that there is a close connection between the orbifold partition function for an inverted oscillator at Euclidean time and the transition amplitude of the normal oscillator at Lorentzian time and similarly for the cases of the inverted oscillator transition amplitude with the normal oscillator partition function. One can then restrict to the study of two out of the four possibilities.

\section{The Grand Canonical Partition Function}
\label{grand}

It is convenient to consider the grand-canonical ensemble instead of the canonical ensemble of the previous section, in order to study the double scaling limit. The canonical partition functions we found in the previous section having the form of determinants and Pfaffians, prove very important in this respect. This is because these forms allow one to pass to the grand-canonical ensemble in a straightforward and rigorous way.
The partition function in the grand canonical ensemble is defined by 
\be
\mathcal{Z}_G= \sum_{N=0}^\infty x^N \mathcal{Z}_N, \qquad x=e^{\beta \mu}\, .
\ee
It is well known in statistical mechanics that typically it is much easier to compute the grand canonical ensemble of fermionic/bosonic
gases rather than the canonical one. In the determinant form this is because one can show that \cite{Kleinert,Marino:2011eh}
\be
\label{gdet}
\mathcal{Z}_G= \sum_{N=0}^\infty \frac{x^N }{N!} \int \prod_{i=1}^N d \lambda_i \det K(\lambda_i, \lambda_j) = \Det \left(I+ x \hat K \right)\, ,
\ee
where $\Det$ is a Fredholm determinant.
The problem is thus reduced to the computation of the spectrum of a Kernel $\hat K$ acting on the space of functions of one variable $f(x)$ as
\be
\hat K \left[f \right](x) = \int dy K(x,y) f(y)\,.
\ee
Therefore now one needs to solve a simpler one-particle problem. A similar equation exists for Pfaffians~\cite{Borodin:2007,Moriyama:2015jsa} 
\begin{align}
\label{gpfaf}
\sum_{n=0}^\infty x^n\int\frac{d^n x}{n!}
(-1)^{\frac{1}{2}(n-1)n}\pf P
=\sqrt{\det\bigl(\overline I-x\, \overline \Omega\, P\bigr)},
\end{align}
where $ P$ is a $2n\times 2n$ skew-symmetric matrix consisting of four $n\times n$ blocks $P_{ab}$ ($a,b=1,2$), whose $(i,j)$-component is $P_{ab}(x_i,x_j)$ satisfying $P_{ba}(x_j,x_j)=-P_{ab}(x_i,x_j)$. The $\overline \Omega$ and $\overline I$ matrices in (\ref{gpfaf}) are defined as, 
\begin{align}
\overline \Omega=\begin{pmatrix}0&I\\-I&0\end{pmatrix},\quad
\overline I=\begin{pmatrix}I&0\\0&I\end{pmatrix}\,.
\label{OmegaI}
\end{align}
Here the Pfaffian on the left-hand side is the finite dimensional one, while the determinant on the right-hand side simultaneously contains a $2\times 2$ determinant and a Fredholm determinant.

\subsection{The Circle}
\label{circlegrand}

The canonical partition function for the circle is of the form~(\ref{gdet}) thus one directly obtains the result
\bea
\mathcal{Z}_G^c = \Det \left(I+ x \hat K \right), \quad  \text{with} \quad K(x,y)&=& \sum_{n=0}^\infty \psi_n (x) \psi_n (y) q^{n+\half},     \nn \\
\text{or} \quad K(z, z')= \frac{1}{q^\half z - q^{-\half} z'}, \quad z&=&e^{i \theta}\,.
\eea
The eigen-functions are either Hermite functions $\psi_n(x)$ or the polynomials $z^n$ with eigenvalues
$\lambda_n = q^{n+\half}$\footnote{One should remember to set $\omega=i$ in case of the inverted harmonic oscillator potential.}.
The partition function and the grand free energy are thus
\bea
\mathcal{Z}_G^c&=& \prod_k \left( 1+ x q^{k+\half} \right)\, , \nn \\
\mathcal{F}_G^c&=& - \sum_k \log \left( 1+ x q^{k+\half} \right) = - \int_{-\infty}^\infty d \epsilon \rho_{H.O.}(\epsilon) \log \left( 1+  e^{\beta_c (\mu - \epsilon)} \right),
\eea
where we introduced the inverted oscillator density of states $\rho_{H.O.}(\epsilon )= \frac{1}{\pi}\sum_k \delta(\e-\e_k)= - \frac{1}{2\pi} Re \Psi(\half + i \e)$, with $\Psi(z)$ the di-gamma function. The derivation of the asymptotic string theory genus expansion from this expression can be found in detail in~\cite{Klebanov:1991qa}.

\subsection{Grand Canonical for the regular representation}
\label{regulargrand}

For the regular representation of the orbifold with $n=N/2$, one can pass to the grand canonical ensemble using the pfaffian formula~\ref{gpfaf}. We also present the $n=0$ case with an alternate method in Appendix~\ref{n0grand}. Combining equation~\ref{pfaffianpf} with~\ref{gpfaf}, the result for the regular representation can be written in a nice operator form as
\be
\mathcal{Z}_G = \sqrt{\det(\overline I+e^{\beta \mu} \widehat{\rho})},
\ee
with 
\begin{align}
\label{rho}
\widehat{ \rho} = \begin{pmatrix}
 \hat{ \mathcal{O}} e^{- \beta \hat H} \hat{ \mathcal{O}} e^{- \beta \hat H} &
 -\hat{ \mathcal{O}} e^{- \beta \hat H} \hat{ \mathcal{O}} e^{- \beta \hat H} \hat{ \mathcal{O}}\\
  -e^{-\beta \hat H} \hat{ \mathcal{O}} e^{-\beta \hat H}  &
e^{-\beta \hat H} \hat{ \mathcal{O}} e^{-\beta \hat H} \hat{ \mathcal{O}}
\end{pmatrix},
\end{align}
where we defined the bi-local operator $\langle x | \hat{\mathcal{O}} | y \rangle = \frac{1}{\pi (x-y)}$ that acts at the orbifold end-points and $\hat{H}$ the usual harmonic oscillator hamiltonian. The evolution is for $\beta= \beta_c/2 $. If we furthermore use the Mehler formula, equation~(\ref{upmehler}), we find that this operator acts on the harmonic oscillator wavefunctions (Hermite functions) at the segment endpoints as
\be
\langle x | \hat{ \mathcal{O}} | \psi_n \rangle =\frac{1}{\pi} \int_{\gamma}d y \frac{\psi_n(y)}{x-y}\,. 
\ee
Now it is important to properly discuss the contour of integration $\gamma$ since the integrand is singular when $x=y$. This is related also to the problem of the singular nature of the integrals we've encountered so far when two eigenvalues of $M_{1,2}$ coalesce. To avoid the singularity one can adopt an $i \e$ prescription to go around the singularity either on the positive or negative imaginary plane, and using the Sokhotski-Plemelj theorem
\be
\frac{1}{x-y \pm i \e} = \mp i \pi  \delta(x-y) + \mathcal{P} \frac{1}{x-y},
\ee 
one learns that these two independent possibilities are either to encircle the singularity and pick a delta function or to adopt the principal value prescription. It is easy to see that the first prescription of the delta function trivialises the action of $\mathcal{\hat{O}}$ and one just finds eigenfunctions of the matrix kernel as the vectors $v^T = (e^{-\beta \hat H} \psi_n , \psi_n )$ and the eigenvalues as $\lambda_n = q^{n+\half}$, $q=e^{- \omega \beta_c}$. The free energy in this case would then just be one for the circle divided by two (due to the pfaffian/square-root of the determinant). \\

This makes clear that the prescription that contains the non-trivial twisted state contribution should be the other one, namely the principal value prescription. In addition, this prescription is consistent with the fact that the original integral is for $y \in (-\infty, \infty)$ and the principal value is the natural regulating prescription for the singular kernel $1/(x-y)$ in this range. One can therefore understand the operator  $\mathcal{\hat{O}}$ acting as a Hilbert transform to the Harmonic oscillator wavefunctions (see appendix \ref{hilbertproperties} for the properties of Hilbert transform.)

\be
\langle x | \hat{ \mathcal{O}} | \psi_n \rangle = \langle x  | \psi^\mathcal{H}_n  \rangle  =\frac{1}{\pi} \mathcal{P} \int_{-\infty}^\infty d y \frac{\psi_n(y)}{x-y} \, .
\ee
One can also notice that the kernel $\hat{\rho}$ can be written as the square of a more elementary kernel $\hat{\rho} = \hat{\bar{\rho}}^2$ with\footnote{Note the similarity with kernels arising in the study of Riemann-Hilbert problems~\cite{Moore:1990cn,Fokas:1991za,Eynard:2015aea}.}
\bea
\langle x | \hat{\bar{\rho}} | y \rangle = \frac{1}{\sqrt{2}} \sum_{n=0}^\infty q^{\half(n+\half)}  \begin{pmatrix}
-\psi_n^\mathcal{H}(x) \psi_n(y)&
\psi_n^\mathcal{H}(x) \psi^\mathcal{H}_n(y)\\
\psi_n(x) \psi_n(y)&
- \psi_n(x) \psi^\mathcal{H}_n(y)
\end{pmatrix}
\label{rhotilde}\,. \nn \\
\eea
One can easily extend these definitions, using from the start parabolic cylinder functions which are the eigenfunctions of the inverse harmonic oscillator and the appropriate Mehler resolution of the propagator, see appendix~\ref{parabolic}. It is also possible then to wick rotate $\beta= i T$ to discuss the real-time propagator as well. It is also important to note that one can also write the orbifold kernel $\hat{\rho}$ in the energy basis in terms of hypergeometric functions, see appendix~\ref{energykernel}. 

Finally, as an interesting result coming from eqn.~\ref{rho}, one can compute the trace of the kernel, if one resolves the operator $ \hat{ \mathcal{O}}$ in momentum basis as $\langle p_1 | \hat{\mathcal{O}} | p_2 \rangle = -i \sgn p_1 \delta (p_1 - p_2)$ (see appendix~\ref{hilbertproperties}). Expressing the oscillator propagator in momentum basis one computes ($\tb = \omega \beta_c$).
\be\label{trace1st}
\tr \hat \rho = \frac{1}{2 \pi \sinh(\tb /2)} \tan^{-1} \frac{1}{\sinh(\tb /2)}\, .
\ee
This is an interesting expression from which we will manage to extract the one-particle density of states -see section~\ref{trkernel1} - and match it with the analogous expression arising from the representation of the kernel in terms of angles that we now turn to.

\subsubsection{Kernel in terms of angles}\label{kangle}

One can find an alternative representation of the kernel in terms of angles using equation~(\ref{anglepf}). To pass to the grand canonical ensemble in this case we used reference~\cite{Borodin:2007} that treats the same structure as we have in terms of angles.
The kernel in this description acts to functions $X(\theta)$ as
\be
 \hat{\rho}\left[ \begin{pmatrix} X_1 \\ X_2 \end{pmatrix} \right] (\theta)=  \int_{0}^{\pi} d \mu(\theta') \rho(\theta, \theta') \begin{pmatrix} X_1(\theta') \\ X_2(\theta') \end{pmatrix},
\ee
with the matrix
\bea\lab{anglematrix}
\rho(\theta, \theta')=\begin{pmatrix} \rho_{11} (\theta, \theta')&\rho_{12} (\theta, \theta')\\ \rho_{21}(\theta, \theta')&\rho_{22}(\theta, \theta')\end{pmatrix}= \begin{pmatrix} \rho_{11}(\theta, \theta') &  \rho_{11}(\theta, -\theta') \\  -\rho_{11}(-\theta, \theta') &  -\rho_{11}(-\theta, -\theta') \end{pmatrix} \nn \\
\rho_{11} (\theta, \theta') = \frac{1}{q^{-1/2} e^{i \theta}-q^{1/2} e^{i \theta'}}+\frac{1}{q^{1/2} e^{-i \theta}-q^{-1/2} e^{-i \theta'}},
\eea
and the measure
\be\lab{anglemeasure}
d \mu(\theta')=\frac{ d\theta'}{2 \pi i}  \frac{q^\half}{\sqrt{(1- q e^{2 i \theta'})(1-q e^{-2 i \theta'})}},
\ee
that contains two branch-cuts in the complex $z'=e^{i \theta'}$ plane, emanating from four points $z'= \pm q^\half , \pm q^{-\half} $. For more details see figs.~\ref{fig:Branchcuts}, \ref{fig:UHP}. The relevant Riemann surface can be understood by gluing two spheres along two branch-cuts, the resulting surface being a torus. In the next section, we see that this kernel simplifies greatly using Jacobi's elliptic functions.

\begin{figure}[t]
\vskip 10pt
\centering
\includegraphics[width=170mm]{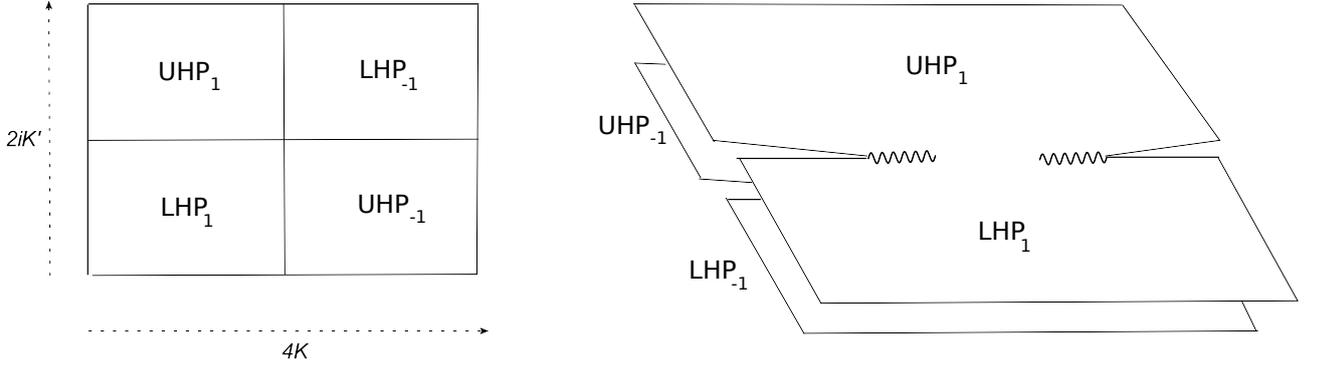}
\caption{The geometry in the complex $z$ plane is of a two-sheeted Riemann surface. The elliptic substitution makes clear that this surface is a torus.}
\label{fig:Branchcuts}
\end{figure}

\subsubsection{Elliptic function parametrization}\label{ellfpar}

We find that the simplest representation of the kernel follows  by going to the double cover and using the doubly periodic elliptic functions. Similar transformations and kernels can be found in studies of Ising, Ashkin-Teller and other models of statistical mechanics \cite{Zamolodchikov:1986,baxter2007,baxter2011onsager}. For more details on elliptic functions the reader can consult~\cite{Lawden}. 
In particular we define $z = e^{i  \theta} = q^\half \sn ( u, q)$ with $\sn u$ Jacobi's elliptic sine. Note that $q \equiv k = e^{- \omega \beta}$ plays the role of the so-called modulus. With this substitution we find
\be
\int_0^\pi d \theta \mu(\theta) \rightarrow - q^\half \int_{K + i K'/2}^{-K + i K'/2} \frac{d u }{2 \pi i}, 
\ee
which is a great simplification for the measure. To find the new range of integration one can follow picture \ref{fig:UHP}. 
\begin{figure}[t]
\vskip 10pt
\centering
\includegraphics[width=176mm]{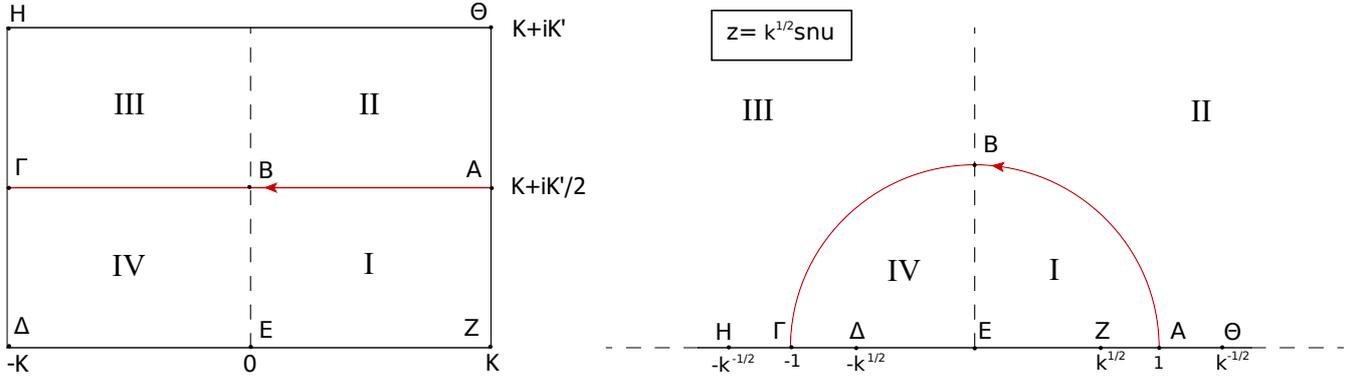}
\caption{The mapping of the rectangle to the upper-half plane via $z= k^\half \sn (u, k)$, with a matching of corresponding points. The branch cuts are between $H \Delta$ and $Z \Theta$. Both pictures correspond to the $UHP_1$ quadrant of \ref{fig:Branchcuts}.}
\label{fig:UHP}
\end{figure}
The eigenvalue equation for the spectrum of the kernel can now be written as follows \small
\be
\lambda \begin{pmatrix} X_1(u) \\ X_2(u) \end{pmatrix}= - q^\half \int_{K + i K'/2}^{-K + i K'/2} \frac{ dv}{2 \pi i}  \begin{pmatrix} \rho_{11}(u,v)&\rho_{11}(u, v + i K')\\ -\rho_{11}(u+ i K',v)&-\rho_{11}(u+ i K', v+ i K')\end{pmatrix} \begin{pmatrix} X_1(v) \\ X_2(v)\,. \end{pmatrix}
\ee
\normalsize
One notices a consistency condition $X_1(u)+X_2(u- i K')=0$ arising from the matrix equation.
We conclude that one need not study a full matrix problem, since the eigenvalue equation reduces to
\bea
\label{elliptickernelfinal}
\lambda X(u) = -q^\half \int_{K + i K'/2}^{-K + i K'/2} \frac{ dv}{2 \pi i} \rho_{11}(u, v)  X(v) - q^\half \int_{-K - i K'/2}^{K - i K'/2} \frac{ dv}{2 \pi i} \rho_{11}(u, v) X(v) \nn \\
=   - q^\half \int_{C_1+C_2} \frac{ dv}{2 \pi i} \rho_{11}(u, v)  X(v), \nn \\
\eea
with
\be
\rho_{11}(u,v) =\frac{1- q \sn u \sn v}{\sn u-q \sn v}\, .
\ee
Let us also note that the Jacobi's sine and thus the kernel, are doubly periodic with periods $4K$, $2iK'$ i.e. $\sn(u+ 4K +2 i K', k) =\sn (u, k)$. 

It is interesting to note that had we instead used the closed contour $C_1+C_2+C_3+C_4$ (see fig.~\ref{fig:closedcontour}), we could have then solved the integral equation by picking the poles of the kernel at $\sn u = q \sn v^*$, finding
\be
\lambda X(u)= q^{-\half} \frac{\cn u}{\cn v^*} X(v^*),
\ee 
which is solved by $X(u)= \cn u \sn^m u , \ m \geq 0$ with eigenvalues $q^{-\half-m}$(if we demand eigenfunctions that are analytic in the interior of the strip of integration/ interior of unit circle), or by $X(u)= \cn u \sn^{-m} u , \ m \geq 1$ with eigenvalues $q^{-\half+m}$ (if we demand eigenfunctions that are analytic in the exterior of the unit circle/strip of integration). This is analogous to the discussion in section~\ref{regulargrand}, where we find an alternative contour that also gives half the free energy on the circle. 

Comparing the integral equation with the contour $C_1 + C_2$ comprising of two horizontal pieces with the one defined via the closed contour $C_1+C_2+C_3+C_4$, we find that we need some extra monodromy data around the torus to relate them. 
This is also to be expected since, the orbifold we consider is more than half of the circle because of the contributions from the twisted states localized at the fixed points of the orbifold.  What we have shown above then means that the information about these twisted states should be contained in the contours $C_3+C_4$. This contribution can be determined either from the contour integrals around the branch cuts or equivalently from monodromy data around the fundamental cycles of the corresponding torus. 

We were not able to solve the integral equation including the contribution from the branch-cuts. Therefore we do not have the full-spectrum of the theory in the $n=N/2$ representation. We list different ways of expressing the Kernel equation in Appendix~\ref{elliptickernel}. These expressions may be useful to obtain the spectrum in future work.

\begin{figure}[t]
\vskip 10pt
\centering
\includegraphics[width=160mm]{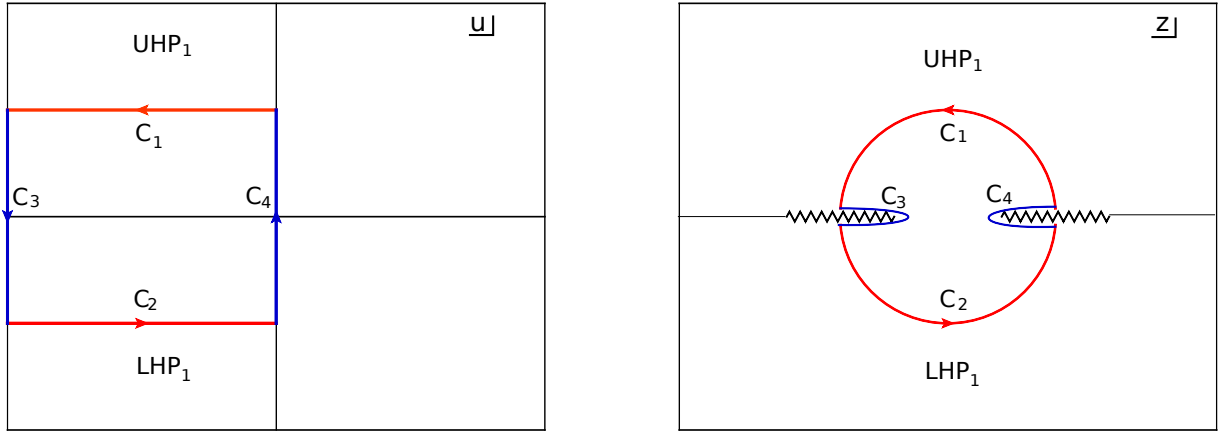}
\caption{The original contour $C_1+C_2$ is drawn by red lines in the $u$ and $z$ plane. In addition we draw also the two extra segments $C_3+C_4$ with which the contour can close. In order to relate the closed with the open contour, one needs to know either the contribution around the torus, or the difference of the integral above and below the branch cut.}
\label{fig:closedcontour}
\end{figure}

\subsubsection{Trace of the kernel}\label{trkernel1}

A consistency check that can be performed in all the different descriptions we have for the kernel is to compute its trace. From~\ref{anglematrix} and~\ref{anglemeasure} we can compute ($\tb = \omega \beta_c$)
\begin{equation}\label{traceangle1}
\tr \hat \rho =\frac{1}{ \sqrt{2} \sinh(\tb /2)} \int_0^\pi \frac{ d \theta}{2 \pi} \frac{\sin{\theta}}{\sqrt{\cosh \tb - \cos(2\theta)}} = \frac{1}{2 \pi \sinh(\tb /2)} \tan^{-1} \frac{1}{\sinh(\tb /2)}\, .
\end{equation}
This equation matches perfectly with eqn.~\ref{trace1st}, derived from the alternative representation of the kernel and thus provides a good consistency check of the two approaches. This equation is to be contrasted with the one-particle oscillator partition function on the circle
\be
Z_{H.O.}^{1p}= \frac{1}{2 \sinh \tb /2}\, .
\ee
By rotating $\omega \rightarrow - i \omega$, one then finds the inverse oscillator result for the orbifold
\begin{equation}
\tr \hat{\rho}_{inv}= \frac{-1}{2 \pi \sin (\omega \beta_c /2)} \tanh^{-1} \left( \frac{1}{\sin (\omega \beta_c /2)} \right)= \int_{0}^{\pi}  \frac{d \theta}{2 \pi} \frac{\cos(\theta/2)}{\cos(\omega \beta_c)- \cos(\theta)}
\end{equation}
for more details see appendix~\ref{trkernel}.
This expression has poles as the usual circle partition function for one particle at $\omega \beta_c = 2 n \pi$, $n \in \mathbb{Z}$ and also branch cuts emanating from $\omega \beta_c =   2\pi(m+1)$, $m \in \mathbb{Z}$  due to the two logarithms from the inverse hyperbolic tangent. One could try to derive the density of states of this partition function using the definition via the Laplace transform
\begin{equation}
\rho_d(\epsilon) =\int_{c -i \infty}^{c + i \infty} \frac{d \beta}{2 \pi i} Z(\beta) e^{\beta \epsilon}\, ,
\end{equation}
but the branch-cuts pose some difficulty. In particular adding the piece at infinity, the contour will enclose all the poles for $n \leq 0$, which gives the same density of states as in the case of the inverted H.O. but in addition one picks contributions from all the branch cuts for $m \leq -1$.  To simplify things, we rewrote this expression as an integral with the integrand having simple poles. Exchanging the integrals, one can formally derive a single particle density of states, $\rho^{1p}_o(\epsilon)= \rho_{H.O.}(\epsilon) + \rho_{twisted}(\e)+\rho_{Im}(\e)$, with the ``twisted" piece
\bea
\rho_{twisted}(\epsilon) &=& \frac{1}{4 \pi \sinh(\frac{\epsilon}{\omega} \pi)} \left[ Im \ \Psi \left( i \frac{\epsilon}{2 \omega} +\frac{1}{4})\right) - Im \ \Psi \left( i \frac{\epsilon}{2\omega} +\frac{3}{4})\right)\right] \nn \\
&=& \frac{1}{4 \pi \sinh(\frac{\epsilon}{\omega} \pi)} Im \ \int_0^\infty dt \frac{e^{-i \frac{\epsilon}{\omega} t}}{\cosh(\frac{t}{2})},
\lab{twist123}
\eea
with $\Psi (z)$ the digamma function.
For more details of this derivation one can see appendix~\ref{trkernel}. Finally, the density of states contains an extra imaginary piece $\rho_{Im}(\e)$ (see~\ref{trkernel}), that might have some interesting interpretation in terms of decaying states, since the decay/tunneling rate of a metastable physical system is related with the imaginary part of the free energy $\Gamma \sim Im \mathcal{F}$~\cite{Affleck:1980ac}.

\section{Large orbifold expansions}\label{lsegment}

The contribution to the partition function from the twisted states can be isolated by considering the limit  $\beta \rightarrow \infty$. This limit reduces the free energy to the ground state contribution as  $\mathcal{F}= \beta_c E_{ground}/2 + \Theta$ and the $\beta$-independent constant piece $\Theta$ in this expression is the twisted state contribution to the ground state energy of the orbifold.  

\subsection{Generic $n$} 

One can obtain a closed form expression for this constant piece in the generic $n$ representation of the orbifold, in the formulation in terms of the eigenvalues of $M$ as follows:
\be\lab{generictheta}
 \Theta = 2 \log \int d^{N_1} x \det_{\begin{subarray}{c} 1 \leq i \leq N \\ 1 \leq k \leq n \\ 1 \leq p \leq N-2n \end{subarray}} \left[ \int dy \frac{\psi_{i-1}(y)}{x_k-y} \quad \int dy y^{N-2n-p} \psi_{i-1}(y) \quad \psi_{i-1}(x_k) \right]
 \ee
In this expression the determinant is of an $N\times N$ matrix with rows labelled by the index $i$ and the  columns separated into three pieces whose size is governed by the range of $k$ and $p$. Derivation of this expression can be found in appendix \ref{genericnlambda}. Another expression in  the second formulation in terms of eigenvalues of $A$ is presented in Appendix \ref{lbgeneric}. We are unable to obtain the analogous expressions after taking the double scaling (large N) limit however. The latter is necessary to make direct connection to the Liouville theory. In principle, one should be able to express $\Theta$ in the grand canonical ensemble. For example it may be possible to obtain it directly using the twisted contribution to the one-particle density of states in equation (\ref{twist123}). In the previous section we also identified the origin of the twisted state contribution through the contour around the branch cuts in figure \ref{fig:closedcontour}. However, none of these alternative formulations has practically helped obtaining the final expression in terms of Liouville theory quantities. Instead, we perform the calculation for specific values of $n$ below.  

\subsection{$n=0$}
 
 Starting from the canonical ensemble, we have managed to treat the $n=0$ representation in terms of even/odd parabolic cylinder functions and write the twisted state contribution in terms of the chemical potential $\mu$.  The relevant calculations are presented with detail in appendix~\ref{lbn0}. This result is found to be
\be 
\Theta = \frac{1}{2}\int^{\mu}\rho_{H.O.} (\e)\int^{\mu}\rho_{H.O.} (\e^{\prime})\log\vert\e -\e^{\prime}\vert d\e d\e^{\prime} .
\ee
One can then use the asymptotic form $\rho_{H.O.} (\e)=\frac{1}{\pi}\le(-\log\e +\sum_{m=1}^{\infty}\mathcal{C}_{m}\e^{-2m}\ri)$ of the density of states to derive the asymptotic genus expansion, which we now describe. In particular one defines the cosmological constant $\Delta = \pi ( \kappa_c^2 - \kappa^2)$ that is related to the \emph{renormalised string coupling} $\mu_0$ as $\Delta = - \mu_0 \log \mu_0$ in the limit $\kappa \rightarrow \kappa_c$. One then fills up states up to the chemical potential $\mu$. The relevant equations are
\be
N=\frac{1}{\hbar} \int^\mu d\e \rho_{H.O.}(\e) \, , \quad \frac{\partial \Delta}{\partial \mu} = \pi \rho(\mu)\, .
\ee
One can invert the second equation above, to find $\mu(\mu_0)$ in an asymptotic expansion whose first term is $\mu = \mu_0$, see~\cite{Klebanov:1991qa}.
After that we can use an asymptotic expansion of the twisted states $\Theta(\mu)$ and turn it into an asymptotic expansion in the renormalised string coupling $\mu_0$.
\bea
\Theta &=&\mu_{0}^{2}\le(\frac{11}{8}-\frac{\pi^{2}}{24}+\le(\frac{\pi^{2}}{12}-\frac{11}{4}\ri)\log{\mu_{0}}+\frac{7}{4}\log^{2}\mu_{0}-\frac{1}{2}\log^{3}\mu_{0}\ri)\nn\\
&&-\frac{1}{24}\le(1+\frac{\pi^{2}}{6}\ri)\log\mu_{0}+\frac{1}{\mu_{0}^{2}}\le(\frac{259}{11520}+\frac{7}{2880}\le(\frac{\pi^{2}}{3}-7\ri)\log\mu_{0}\ri)\mathcal{O}(\mu_{0}^{-4}), \nn \\
\eea
with $\mu_0$ the renormalised string coupling. One notices that the torus contribution is not the same as in equation~\ref{orblfinal}.

\subsection{$n=N/2$}

We will now finally treat the case that provides the matching between Liouville theory and the matrix model.
For the regular representation $n=N/2$ the generic expression in (\ref{generictheta}) simplifies as 
\be\lab{twN2}
\Theta= \half \log \det O_{ij}, \qquad  O_{ij} = 2 \int_{-\infty}^\infty dx dy \frac{\psi^+(\e_i, x)\psi^-(\e_j, y)}{x-y}  \, .
\ee
We have calculated this expression using both the Hermite functions and the delta-function normalised even and odd parabolic cylinder functions which are the eigenfunctions of the inverted oscillator, see appendix~\ref{parabolic}. Details of the  calculation  of (\ref{twN2}) are presented in appendices \ref{lbregular} and~\ref{lbregularparabolic} respectively. For the normal oscillator one finds that the result can be expressed in terms of a determinant of \emph{sine-kernel}, see~ \ref{sinekernelnormal}. For the inverted oscillator the result is in terms of continuous labels 
\be\lab{res2}
O(\e_1,\e_2) = \frac{1}{\pi} \lvert \Gamma(1/4+ i \e_1/2) \Gamma(3/4+ i \e_2/2)\rvert  \e^{\pi (3 \e_2 + \e_1)/4}   \frac{ \sinh\left(\frac{1}{4} \pi
   (\e_2-\e_1)\right)}{\e_1-\e_2} \, .
 \ee
Substituting this in (\ref{twN2}), we see that the determinant becomes the product of diagonal pieces times the determinant of the $\sinh(\e_1-\e_2)/(\e_1-\e_2)$. It is easy to see that the diagonal pieces do not contribute to the $1/\mu$ expansion in the double scaling limit, only giving contributions to non-perturbative terms in $\mu$. Therefore the twisted state contribution to the perturbative expansion in $g_s = 1/\mu_0$ is determined by the kernel of the operator $\sinh(\e_1-\e_2)/(\e_1-\e_2)$ in the double-scaling limit. We should also remember to solve $\mu(\mu_0)$ to derive the correct asymptotic expansion.

It is, as far as we know, not possible to calculate the spectra of this kernel with the currently available methods. However the determinant of sine kernel where one replaces $\sinh(\e_1-\e_2)$ with $\sin(\e_1-\e_2)$ is possible to be calculated in an asymptotic fashion as was done in the 70s~\cite{Dyson:1976nq,Deift,Krasovsky}. Luckily, we can make the replacement $\epsilon \to i \epsilon$ in (\ref{res2}), hence transform the $\sinh(\Delta \epsilon)/\Delta \epsilon$  kernel  into   $\sin(\Delta \epsilon)/\Delta \epsilon$ kernel by considering the following, alternative calculation. The canonical partition function (\ref{eigenvaluerepn}) with the propagator (\ref{upmehler}) is invariant under 
$\omega \to i \omega$, $\beta\to -i \beta$. 

This is because one can Wick rotate the integrals over the matrix eigenvalues as $x_i \to e^{-i \pi/4} x_i$, $y_i \to e^{-i \pi/4} y_i$ in the partition function. To see this consider the integral along the contour $C = (-\infty, \to \infty)  \cup (\infty, \infty e^{-i\pi/4}) \cup ( \infty e^{-i\pi/4},  -\infty e^{-i\pi/4}) \cup (-\infty e^{-i\pi/4}, -\infty)$ where the second and the last pieces are on the indicated arcs at infinity.  One can see that there are no poles inside this contour $C$ as follows. The only possible poles could arise from the denominator in the initial and final wave functions in (\ref{inifin}). However these poles can easily be avoided by rotating $x$s and $x'$s (and similarly $y$s and $y'$s) in pairs. Also,  there are no possible divergences at the arcs at infinity, $|x|=\infty$, in the $n=N/2$ partition function we are interested in here because the wave functions, (\ref{inifin}) decay at infinity in this case\footnote{Note that this part of the argument would fail for the partition functions with $n<N/2$.}. Finally, one shows that possible divergence that could arise from the $\det K(\bar{x},\bar{x}')$ in (\ref{eigenvaluerepn}) on the infinite arcs in contour $C$ are also absent because one can expand
\be\lab{expK} 
\det K = \sum_{r} \bar{\Psi}_r(\bar{x}) \Psi_r(\bar{x}') e^{i \beta E_r}\, 
\ee 
 where the N-fermion wave functions $\Psi_r$ are constructed out of products of the the parabolic cylinder wave-functions, and the latter are convergent on the particular infinite arcs $ (\infty, \infty e^{-i\pi/4}) $ and $ (-\infty e^{-i\pi/4}, -\infty)$, as can be seen from appendix~\ref{parabolic}. We conclude that the integral on the contour $C$ vanishes, thus one can Wick-rotate $x_i \to e^{-i \pi/4} x_i$, $y_i \to e^{-i \pi/4} y_i$ in the partition function giving rise to the symmetry under $\omega \to i \omega$, $\beta\to -i \beta$. 
 
Thus, one could calculate the twisted state contribution in the Lorentzian path integral instead of the Euclidean partition function. The only difference that this makes for the twisted state contribution coming from (\ref{res2}) is to replace the energies $\epsilon \to i \epsilon$, thus transforming\footnote{One may be ask how come the twisted state contributions in the Euclidean and Lorentzian path integrals give rise to different expressions. After all twisted states that are localized on the fixed points are not supposed to see the signature of time. This should be true at the non-perturbative level. Asymptotic expansions can differ, which is the well known Stokes phenomenon.}  the sinh into the sine kernel. This is also the result for the normal oscillator.

Therefore the result, remarkably, boils down to the computation of a Fredholm-determinant of \emph{sine-kernel} which is a well known object in random matrix theory~\cite{Mehta} that corresponds to the probability that all the energy eigenvalues are outside the energy range $(-\mu, 0)$ and thus form the fermi sea.
This object has been computed with various approaches such as inverse scattering, toeplitz determinants and the Riemann-Hilbert method. Some basic references are~\cite{Dyson:1976nq,Deift,Krasovsky}. This calculation is reviewed in appendix~\ref{LevelSpa} and results in 
\be 
\Theta =\frac{1}{4} \log E_2 (0 ; (0,\mu_0)) =  - \frac{1}{32} \mu_0^2 - \frac{1}{16} \log \mu_0 +\frac{1}{48}\log 2 + \frac{3}{4}\zeta'(-1) + O\left( \frac{1}{\mu_0^{2m}}\right) \, . 
\ee
We observe that the twisted state contribution to the torus level partition function $- \frac{1}{16} \log \mu_0$ matches precisely the world-sheet result (\ref{orblfinal}). This provides a non-trivial check of the duality we propose between the $n=N/2$ representation of the orbifold matrix quantum mechanics and the 2D non-critical string theory on $S_1/\mathbb{Z}_2$.  

%
%

\section{Conclusions}\label{conclusions}

In this paper we considered the quantum mechanics of an $N\times N$ dimensional Hermitean matrix  $M$ compactified on Euclidean time $\tau$ and orbifolded by a $\mathbb{Z}_2$ action that contains the reflection $\tau\to -\tau$, which we also embedded into the gauge group. 
We provided evidence that this MQM on the $S^1/\mathbb{Z}_2$ orbifold in the large-N limit constitutes a good toy model for a Bang-Crunch universe in the context of 2D string theory. This is because the orbifold MQM admits a natural analytic continuation into Lorentzian time as shown in equation (\ref{decomp}) and in the double scaling limit the theory becomes dual to 2D string theory with space-like singularities at Lorentzian time $t=0$ and $t=T$. The space-like dimension of this 2D string theory is given by the Liouville direction that is made out of the eigenvalues of $M$ in the dual MQM description.

\paragraph{Partition function -}
The information that one can practically extract from the Liouville description of this theory is rather limited at the moment. In particular we managed to compute the torus contribution to the partition function including the contribution of the twisted states by indirect consistency methods as shown in section \ref{string}. On the other hand, we believe that the description of the theory in terms of MQM provides an alternative, richer point of view.

As a first step, we focused on calculating the partition function of the orbifolded MQM. We found that the orbifolding operation in the MQM description can be given in terms of different representations labeled by a parameter $0 \leq n \leq N/2$ (with even N). These representations  arise from possible embeddings of $\mathbb{Z}_2$ into the $SU(N)$ gauge group. We argued why the ``regular'' representation with $n=N/2$ is preferred. We also showed that the different representations are connected by the action at the orbifold fixed points of operators resembling loop operators in section \ref{loop}. These operators should correspond to changing the number of stretched open strings between the two sets of $n$, $N-n$ $D0$ branes. 

We calculated both the canonical and the grand canonical partition functions using two different formulations. The first formulation involves first integrating over the gauge field and represents the partition function as an integral over the eigenvalues $\lambda_i$ of the matrix $M$. The final expression for an arbitrary representation $n$ is given in equation (\ref{eigenvaluerepn}). 
This representation is useful since as we show in equation (\ref{decomp}) the integrand can be naturally decomposed into a piece localized at $\tau=0$, a transition amplitude from $\tau=0$ to $\tau=\beta$ and a piece localized at $\tau=\beta$. This form of the partition function therefore admits a natural rotation into Lorenzian time where the first and the last pieces are naturally identified with the initial and final wave-functions of the toy cosmological universe, and the middle piece with the transition amplitude from the big-bang to the big-crunch. These wave functions depend on the orbifold index $n$, hence in some sense provide us with a classification of possible bang/crunch universes in this toy model and hence it is crucial to understand the role of $n$ from the string theory side as well. We also note a similarity of our wavefunctions with the ones arising in the work of Dijkgraaf/Vafa on ``negative branes'' and supermatrix models, see~\cite{Vafa:2014iua}. We do not develop these observations further in this paper. One should be really careful about whether the Wick rotation into Lorentzian time applies smoothly near the singularities/end points in time. Finally, there is always the possibility of inserting excited states at the initial and final states of the universe. Nevertheless, this description suggests an intriguing general qualitative prescription for how to make sense of quantum gravity  in  a bang/crunch cosmology: express the theory in terms of a dual open-string description, evaluate the orbifold partition function in Euclidean time to obtain a decomposition into pieces that contain the initial state, transition and the final state, and finally Wick rotate into Lorentzian time.

The second formulation of the partition function involves first integrating over the matrix $M$ and expressing the result in terms of the eigenvalues of the gauge field $A$. This method gives an alternative form for the partition function in terms of Wilson lines, the zero modes of the gauge field. The final expression for an arbitrary representation $n$ is given in equation (\ref{ncanonical}). This formulation clarifies the meaning of the index $n$: as shown in (\ref{wilsonorb}), in the T-dual picture, $n$ corresponds to the number of free D-instantons -free to move along the time direction. There also exists $N-2n$ fractional D-instantons stuck at the fixed points of the orbifold. Thus there are no fractional D-instantons in the regular representation with $n=N/2$ and there are only fractional instantons in the $n=0$ representation.

The $n=N/2$ partition function in this formulation contains a measure which can be thought as containing vortex/Wilson line perturbations of arbitrary order in the form of $\exp \sum_k t_k \left( \tr U^k + \tr U^{-k} \right)$ with $t_k= q^k/2k$. Similar deformations are encountered also in versions of the GWW model~\cite{Gross:1980he,Jurkiewicz:1982iz,Periwal:1990gf} which has a third order phase transition, as well as in the proposed matrix model description of the $SL(2,\mathbb{R})/U(1)$ 2D black hole~\cite{Kazakov:2000pm}. A possible issue with that proposal is that it is based on the FZZ correspondence with the Sine-Liouville which holds for the radius $R=\frac{3}{2}$ close to the black hole-string correspondence point~\cite{Giveon:2005mi}. In contrast, in our case these deformations include all windings and are temperature or radius dependent, which is a quite interesting novel characteristic. In addition it is expected that the higher-windings we find are related to higher spin generalisations of the 2D black hole~\cite{Mukherji:1991kz}, where discrete states are liberated as well~\cite{Mukherjee:2006zv}.
These discrete states are remnants of the higher-spin excitations that exist in higher dimensions~\cite{Ginsparg:1993is,Martinec:2004td} and it is not unnatural to expect their presence due to the orbifolding and breaking of the gauge group that liberates $SU(N)$ non-singlet states near the end of time. The closed string twisted states should then be thought of as a condensate of both the tachyon and those extra states. The possible presence of these states due to the temperature dependent higher winding perturbations can thus lead to quite interesting and rich physics once we manage to compute the partition function or other observables for finite orbifold size $R$.

The two formulations should of course be equivalent. Even though we have not managed to find a direct change of variables that would relate the two in the canonical ensemble, the equivalence can be partially demonstrated at the level of the grand canonical ensemble. Indeed, in both formulations it is possible to go to the grand canonical ensemble and express it in terms of a square root of a Fredholm determinant of a one-particle kernel $\hat \rho$. The spectrum of this Kernel then determines the full non-perturbative answer. We checked the 
equivalence of the two formulations by explicitly matching the trace of this Kernel in the two cases, see equations (\ref{trace1st}) and (\ref{traceangle1}). 

\paragraph{Twisted states -}

A central focus of our paper is the contribution of the twisted states to the orbifold partition function. Since these states are localized at the fixed points of the orbifold that are supposed to become the cosmological singularities under Wick rotation, they are expected to contain crucial information on the string dynamics around these points. The twisted states are clearly marked in the torus partition function of the Liouville theory. Their contribution is given by the constant ($R$-independent) terms in section \ref{string}. One can isolate this contribution in the dual MQM partition function in the first formulation (in terms of eigenvalues of $M$) by taking the large $\beta = \pi R$ limit. This limit, essentially decouples the propagation from the wavefunctions/states at the endpoints in time and focuses on the ground state channel contribution to the free energy. The radius independent piece has the form of determinant operators and was denoted by $\Theta$ in section \ref{lsegment}. We were able to explicitly express and compute $\Theta$ in terms of 2D string theory parameters in the $n=0$  and $n=N/2$ representations. This provided the exact matching with the Liouville theory prediction for the torus for the $n=N/2$ representation. 

It is also interesting to single out the twisted state contribution directly at the level of the grand canonical ensemble. In particular we worked out the regular $n=N/2$ representation and found that they should manifest in the spectrum of the one particle kernel $\hat{\rho}$ which can be determined solving an integral equation. In the first formulation, section~\ref{regulargrand}, the presence of extra twisted states was understood through the action of \emph{Hilbert transform operators} at the endpoints. The large $\beta$ limit, again decouples the Hamiltonian propagation from these operators and ``zooms in'' at the endpoints in time.

In the second formulation (in terms of eigenvalues of $A$), we isolated this contribution in section \ref{ellfpar}. Here the integral is defined on a complex plane with two branch cuts, or alternatively on a two-torus. One obtains precisely half the free energy for MQM on $S^1$ if one ignores the contribution to the contour of integration around these branch cuts, or alternatively the monodromy around the fundamental cycles of the corresponding torus. Hence in this description the twisted states should be contained in these branch-cut or monodromy contributions. Moreover, let us note that from the Matrix model picture it is clear that these extra contributions can generically lead to both radius dependent together with radius independent terms in the free energy.  
 
We also note that in both formulations, the partition function looks very similar to a four point correlation function: in the first formulation it can be thought of as a correlator between two bi-local operators and in the second as containing four twist operators creating the two branch cuts.

\paragraph{Future directions -}

In this paper we focused on the closed string asymptotic expansion of the partition function. We have found that the matrix model also contains a wealth of non-perturbative information.

It will be interesting to understand further the contribution of the fractional instantons present in other representations, which we expect to be non-perturbative in $g_{st}$.

Let us also note that the structure of the partition function in terms of Wilson lines, is very reminiscent of $\tau$ functions of $BKP/DKP$ Hierarchies~\cite{Jimbo:1983if,vandeLeur:2015,Orlov:2016} and it may be very interesting to pursue this connection. For further progress in this direction, one should study free fermions and $\tau$-functions in the presence of twist fields.

Some other interesting calculations we look forward to perform in the future include the disk one point function and the annulus correlation function for two macroscopic loops. Such quantities will be very good probes of the singularities at the endpoints of time.

Furthermore, we should develop the target space picture of our construction further by using the relation between the matrix eigenvalues and the Liouville coordinate $\phi$. A description of the initial state in collective field theory variables might prove useful here. A natural question in this context is, what is the spatial extend of the 2D universe near the singularities? Is our theory describing one of the known metrics in the 2D string theory literature? The previous probes we mentioned could also help in giving answers to these questions.

As a final observation we recall that~\cite{Dijkgraaf:1991ba,Giveon:1994fu} the horizon and the singularity of the 2D black hole is exchanged under T-duality and that there is a relation between the 2D cosmology with the 2D black hole~\cite{Tseytlin:1991xk}. This can be shown at least at the classical level, for the  Lorentzian 2D black hole~\cite{Witten:1991yr,Mandal:1991tz} described by the $SL(2,\mathbb{R})/U(1)$ WZW coset. It is interesting to note that the Hilbert transform operators at the endpoints in time commute with the $SL(2,\mathbb{R})$ generators of linear fractional transformations and that the description of the kernel on the torus has a manifest $SL(2, \mathbb{Z})$ symmetry. In addition, based on the fact that we have a combination of radius dependent vortex perturbations together with radius independent twisted states, it would be very interesting to investigate whether we can similarly relate our setup with a 2D black hole with a possible interpretation of the twisted states as black hole microstates. To this end, it is encouraging that the contribution of the end-point wavefunctions to the canonical free energy takes the form of an entropy $S\sim \tr log \rho_{twisted}$ (or $S= N \log 2$ for the normal oscillator), which is also the logarithm of the probability of forming the fermi-sea from an ensemble of random hermitean hamiltonians (taking the double scaling limit of the inverted oscillator). 
For all these reasons it would be extremely interesting to investigate similar $S^1/\mathbb{Z}_2$ orbifolds in higher dimensions\footnote{In the context of the 4D Schwarzschild black hole analogous \emph{Lorentzian} $Z_2$ involutions that involve time reversal have found a recent interest in~\cite{Betzios:2016yaq}.}.


\section*{Acknowledgements}

This paper is a continuation of the unpublished work one of the authors (U.G.) has done in collaboration with Hong Liu in 2005. Therefore the basic idea and some of the results have been obtained in collaboration with Hong Liu and we are grateful to his collaboration at the early stages. We also wish to thank Marcos Crichigno, Vladimir Kazakov, Alexei Morozov and Elli Pomoni for useful conversations.
Finally we happily acknowledge discussions with Johan van de Leur and Alexander Orlov in relation to integrable hierarchies and Alexander R. Its on the Riemann-Hilbert problems.

This work is supported by the Netherlands Organisation for Scientific Research (NWO) under
the VIDI grant 680-47-518, and the Delta-Institute for Theoretical Physics (D-ITP) that is funded by the Dutch Ministry of Education, Culture and Science
(OCW).

\appendix

\section{Other classes of orbifolds}\label{susyorb}
Here we present the rest of the supersymmetric orbifold theories for completeness.
\item \textbf{Orbifold II}: The second class of orbifolds are obtained by modding out super-affine theories by
the same reflection symmetry as above. One has the following relations \cite{Dixon:1988ac}:
\be 
\mathcal{Z}_{orb−saA,B}(R)=\frac{1}{2}\mathcal{Z}_{saA,B}(R)+const_{A,B}
\ee
and the following relation at the special radius \cite{Dixon:1988ac}:
\be 
\mathcal{Z}_{orb−saA,B}(\sqrt{2})=\mathcal{Z}_{cirA,B}(\sqrt{2})\,.
\ee
The partition functions are:
\be
\mathcal{Z}_{orb−saA}(R)=\frac{1}{2}\mathcal{Z}_{saA}(R)-\frac{1}{8}\ln\mu_0 ,\,\, \mathcal{Z}_{orb−saB}(R)=\frac{1}{2}\mathcal{Z}_{saB}(R)-\frac{1}{16}\ln\mu_0
\ee
These theories are seperately self dual under $R\rightarrow 2/R$.
\item \textbf{Orbifold III}: The third class of orbifolds are obtained by twisting the circular theories by $(−1)^{F_s}R$.
Note that this is only a symmetry in the 0A theory. One obtains,
\be
\mathcal{Z}_{orbA}(R)=\frac{1}{2}\mathcal{Z}_{cirA}(R)+const
\ee
and
\be 
\mathcal{Z}_{orbA}(1)=\mathcal{Z}_{saA}(2)
\ee
The result is:
\be
\mathcal{Z}_{orbA}(R)=\frac{1}{2}\mathcal{Z}_{cirA}(R)-\frac{1}{8 \sqrt{2}}\ln\mu_0
\ee
We observe that orbA and orbB theories are exchanged under T-duality.
\end{itemize}

\section{Oscillator wavefunctions}\label{parabolic}
We provide this section as a collection of the relations between various representations of normal/inverted harmonic oscillator wavefunctions.

\subsection{Normal Harmonic Oscilator}
We define the normal harmonic oscillator time independent Schroendinger equation $\hbar, m =1$
\be
\half \left( -\partial_x^2 +\omega^2 x^2 \right) \psi_n = \e_n \psi_n\,.
\ee
The Kronecker delta normalised eigenfunctions are
\be
\psi_n(x) = \frac{1}{\sqrt{2^n n!}} \left(\frac{\omega}{\pi}\right)^{\frac{1}{4}} e^{- \half \omega x^2} H_n(\sqrt{\omega} x) =  \frac{1}{\sqrt{n!}} \left(\frac{\omega}{\pi}\right)^{\frac{1}{4}} D_n(\sqrt{ 2 \omega} x)\,.
\ee
These wavefunctions satisfy Mehler's formula for the propagator (in real time)
\be
\left(\frac{\omega}{2\pi i \sin \omega T} \right)^{\half} e^{\frac{i \omega}{2 \sin \omega T}\left[\left( \lambda_i^2 +  {\lambda'}_{j}^{2} \right) \cos \omega T - 2  \lambda_i \lambda'_{j}  \right] } \nn \\
= \sum_n \psi_n (\lambda_i) \psi_n (\lambda'_{j}) e^{- i \omega (n+\half) T},
\ee
which we analytically continued to Euclidean time via $T=-i \beta$ to obtain equation~\ref{upmehler}.

\subsection{Inverted Harmonic Oscillator}

We want to solve the inverse harmonic oscillator time independent Schroendinger equation. Take the normal harmonic oscillator equation and let $\omega \rightarrow i$, and $x \rightarrow x / \sqrt{2} $. One then needs to solve:
\be
\label{inverted}
\left( \partial_x^2 + \frac{x^2}{4} \right) \psi = \e \psi\,.
\ee
This is a particular form of the Weber differential equation
\be
\left( \partial_z^2 +\nu +\half - \frac{z^2}{4} \right) \psi =0\, .
\ee
The solutions are the Parabolic cylinder  functions (equivalently expressed via Whittaker functions W)
\be
D_{\nu}(z) =2^{\frac{\nu}{2}+\frac{1}{4}} z^{-\half} W_{ \frac{\nu}{2}+\frac{1}{4}, -\frac{1}{4}}( \frac{z^2}{2}), \quad  D_{-\nu - 1}(i z)= 2^{\frac{-\nu}{2}-\frac{1}{4}} e^{i \pi/4} z^{-\half} W_{ \frac{-\nu}{2}-\frac{1}{4}, -\frac{1}{4}}( -\frac{z^2}{2})\,.
\ee
where $D_\nu(z), D_{-\nu -1}(\pm i z)$ are linearly independent.
 We are in the specific case where $\nu=i \e-\half$, $i x^2=  z^2$, thus
\be
D_{i \e - \half}(e^{i \frac{\pi}{4}} x) =\frac{2^{\frac{i \e}{2}} e^{- i \pi /8}}{ x^{\half}} W_{ \frac{i \e}{2}, -\frac{1}{4}}( \frac{i x^2}{2}), \quad  D_{-i \e - \half}(e^{i \frac{3\pi}{4}} x)= \frac{2^{\frac{-i \e}{2}} e^{i \frac{3 \pi}{8}}}{ x^{\half}} W_{ \frac{-i \e}{2}, -\frac{1}{4}}( - i \frac{x^2}{2}),
\ee
are the two linearly independent solutions in our case and there is a degeneracy in the continuous energy spectrum. It is easy to see that they are also formally obtainable from the normal harmonic oscillator upon substituting $x \rightarrow x/\sqrt{2}$, $\omega= \pm i$ and $n=\pm i \e - \half$, the normalization is different though. \\

Another useful basis of solutions are the delta function normalised even/odd parabolic cylinder functions~\cite{Moore:1991sf} which we will denote by $\psi^\pm (\e, z)$
\bea
\psi^+(\e, x) = \left(\frac{1}{4 \pi \sqrt{(1+e^{2 \pi \e})}}\right)^\half 2^{1/4} \lvert \frac{\Gamma(1/4+ i \e/2)}{\Gamma(3/4+ i \e/2)} \rvert e^{- i x^2/4} {_1F_1} (1/4-i \e /2 , 1/2 ; i x^2/2) \nn \\
=\frac{e^{- i \pi/8}}{2 \pi} e^{- \e \pi /4} |\Gamma(1/4 + i \e/2)| \frac{1}{\sqrt{|x|}} M_{i \e/2 , - 1/4}(ix^2/2) \nn \\
\psi^-(\e, x) = \left(\frac{1}{4 \pi \sqrt{(1+e^{2 \pi \e})}}\right)^\half  2^{3/4}  \lvert \frac{\Gamma(3/4+ i \e/2)}{\Gamma(1/4+ i \e/2)} \rvert x e^{- i x^2/4} {_1F_1} (3/4-i \e /2 , 3/2 ; i x^2/2) \nn \\
=\frac{e^{-3 i \pi/8}}{ \pi} e^{- \e \pi /4} |\Gamma(3/4 + i \e/2)| \frac{x}{|x|^{3/2}} M_{i \e/2 ,  1/4}(ix^2/2)\,. \nn \\
\eea
Their normalisation is
\be
\int_{-\infty}^\infty dx \sum_{a=\pm} \psi^a (\e_1 , x) \psi^a (\e_2 , x) = \delta(\e_1 - \e_2),
\ee
and
\be
\int_{-\infty}^\infty d\e \sum_{a=\pm} \psi^a (\e , x_1) \psi^a (\e , x_2) = \delta(x_1 - x_2)\,.
\ee
The relation with the previous basis can be established using the following equations
\bea
D_{ i \e -\half}(e^{i \pi/4} x)= \frac{\sqrt{\pi} 2^{i \e/2} e^{-i \pi/8}}{\Gamma(3/4-i\e/2) \sqrt{x}} M_{i \e/2 , -1/4}(i x^2/2)
- \frac{2 \sqrt{\pi} 2^{i \e/2} e^{-i \pi/8}}{\Gamma(1/4-i\e/2) \sqrt{x}} M_{i \e/2 , 1/4}(i x^2/2)  \nn \\
D_{- i \e -\half}(e^{i 3\pi/4} x)= \frac{\Gamma(1/2-i\e)}{\sqrt{2 \pi}} \left[e^{-\e \pi/2} e^{i \pi/4} D_{ i \e -\half}(e^{i \pi/4} x) + e^{\e \pi/2} e^{-i \pi/4} D_{ i \e -\half}(-e^{i \pi/4} x) \right] \,.\nn \\
\eea

\subsection{Mehler for parabolic cylinder}\label{Mehler} 
The delta-function normalised odd/even parabolic cylinder functions $\psi^\mp (\e , x)$ satisfy the following formula~\cite{Moore:1991sf}:
\be
\langle x | e^{-2 i T H}| y \rangle= \int_{-\infty}^\infty d \e e^{i \e T} \sum_{a=\pm} \psi^a(\e , x) \psi^a(\e , y) = \frac{1}{\sqrt{4 \pi i \sinh T}} \exp \frac{i}{4} \left[\frac{x^2 + y^2}{\tanh T} - \frac{2 x y}{\sinh T} \right],
\ee
which is the analogue of Mehler's formula for the real-time ($T=- i \beta$) inverted H.O. propagator with the Hamiltonian~\ref{inverted}. This holds for $-\pi< Im T < 0$ or $Im T=0$ with $Re T \neq 0$. To prove it one can use the general expression (7.694) in~\cite{Jeffrey:2007}. \\
An equivalent expression can be found also in the basis of $D_{i \e -\half}(z), D_{-i \e -\half}(i z)$ using (7.77.3) of~\cite{Jeffrey:2007}
\bea
\langle x | e^{-2 i T H}| y \rangle  =  \frac{1}{\sqrt{4 \pi i \sinh T}} \exp \frac{i}{4} \left[\frac{x^2 + y^2}{\tanh T} - \frac{2 x y}{\sinh T} \right] \nn \\
= \int_{- \infty}^\infty d \e e^{i \e T} \frac{e^{-\half \e \pi}}{4 \pi \cosh(\e \pi)} \left[D_{i \e - \half}(e^{i \frac{\pi}{4}} x) D_{-i \e - \half}(e^{i \frac{3\pi}{4}} y)  + D_{i \e - \half}(-e^{i \frac{\pi}{4}} x) D_{-i \e - \half}(e^{i \frac{3\pi}{4}} y)  \right], \nn \\
\eea
with the same restrictions in T. These expressions are most well suited to compute the transition amplitude in real time. To recover the Euclidean, inverted H.O. expression one needs to set $T=- i \beta$ in the above, (note that they will hold for $R<1$, otherwise one needs to change the contour of integration to make them well behaved).

\section{Representation in terms of angles (Wilson-lines)}\label{anglerep}
Instead of integrating out $U$ one can first integrate out the $M$'s in the expression
\be
\mathcal{Z}_{n, N} = \int \mathcal{D}M \mathcal{D}M' \mathcal{D}U \langle U M' U^\dagger, \beta | M, 0 \rangle = \int \mathcal{D}U I(U)\,.
\ee
If we define $A=1/\tanh(\omega \beta), \ B= 1/\sinh(\omega \beta)$, and remember to use blocks for the matrices after orbifolding we get
\bea
&I(U)=\omega^{-\half (N-2n)^2} \left(\frac{B}{2 \pi}\right)^{N^2/2} \int dM_1 dM_2 dM_1' dM_2' e^T, \quad  U= \begin{pmatrix}
U_1 & U_{12} \\ U_{21} & U_2
\end{pmatrix}, \nonumber \\ 
&K= -\frac{A}{2} \mathrm{tr} (M_1^2+ M_1^2{}') +B \mathrm{tr}(M_1 U_1 M_1' U_1^\dagger+ M_1 U_{12}M_2' U_{12}^\dagger) + (1\leftrightarrow2)\,.
\eea
Now the $U's$ are complex but satisfy certain conditions 
\bea\label{relations}
U_1 U_1^\dagger + U_{12}U_{12}^\dagger = U_2 U_2^\dagger + U_{21}U_{21}^\dagger =1 , \quad U_{12} U_{12}^\dagger = U_{21} U_{21}^\dagger \\
U_1 U_{21}^\dagger + U_{12}U_2^\dagger = U_2^\dagger U_{21} + U_{1 2}^\dagger U_1 = 0,
\eea
and can be diagonalised by bi-unitary transformations such that they leave the measure invariant. We thus use the unitary matrices $V_1, V_1', V_2 , V_2'$ to get\footnote{Note that any complex matrix can diagonalized by bi-unitary transformations. Also the first line of equation~\ref{relations} implies that $U_1 U^\dagger_1$ and $U_{12} U^\dagger_{12}$ can be simultaneously diagonalized.}
\bea
U_1= V_1 C V_1'^\dagger, \quad U_2=V_2 \begin{pmatrix} C & 0 \\ 0 
& 1 \end{pmatrix} V_2'^\dagger, \quad U_{1 2}= - V_1 (D, 0) V_2'^\dagger, \quad U_{2 1}= V_2 \begin{pmatrix}
D \\ 0
\end{pmatrix} V_1^\dagger ,\nn\\
\eea
with
\be
C_{i j}= \cos \theta_i \delta_{i j} , \quad D_{i j}= \sin \theta_i \delta_{i j}, \quad 0\leq \theta_i \leq \frac{\pi}{2}\,.
\ee
This can be also easily achieved after exponentiation of the zero mode of $A$ that has only non-zero the diagonal components of the off-diagonal blocks.\\
Since the measure of M's is invariant under a unitary transformation, we can write the four matrix coupling term of K as
\be
\mathrm{tr}(M_1 C M'_1 C + M_1 D R D + R D M'_1 D + R C R' C + S S'^\dagger C + S^\dagger C S' + T T')
\ee
where we have written $M_2$ and $M'_2$ (which are $(N - n)\times (N - n)$ matrices) as
\be
M_2= \begin{pmatrix} R & S \\ S^\dagger & T \end{pmatrix}, \quad M'_2= \begin{pmatrix} R' & S' \\ S'^\dagger & T' \end{pmatrix}
\ee
with $R, R'$ $n \times n$ matrices. Integration over $T, T'$ will yield a constant factor
\be
(2 \pi)^{(N- 2n)^2}
\ee
Integrations over $S, S'$ yield
\be
(2 \pi)^{2n(N- 2n)} \prod_{i}^n\left(\frac{B^2}{1+B^2 \sin^2 \theta_i} \right)^{N-2n} 
\ee
and integrations over $R, M_1$ give
\be
(2 \pi)^{2n^2} \prod_{i,j}\left( \frac{1}{(1+B^2 \sin^2 (\theta_i+\theta_j) (1+B^2 \sin^2( \theta_i-\theta_j)} \right)^{\half}
\ee
Thus altogether we get 
\be
I= \left(\frac{2 \pi B}{\omega}\right)^{\frac{ (N-2n)^2}{2}} \prod_{i}^n\left[\frac{B^2}{1+B^2 \sin^2 \theta_i} \right]^{N-2n} \prod_{i,j}^n\left[ \frac{B^4}{(1+B^2 \sin^2 (\theta_i+\theta_j) (1+B^2 \sin^2( \theta_i-\theta_j)} \right]^{\half}
\ee 
It is also useful to massage this expression into
\be
\label{I}
I= \left[\frac{2 \pi B}{\omega}\right]^{\frac{ (N-2n)^2}{2}} \prod_{i}^n\left[\frac{2}{\cosh  \tb - \cos  \theta_i} \right]^{N-2n} \prod_{i,j}\left[ \frac{4}{(\cosh \tb - \cos (\theta_i+\theta_j) (\cosh \tb -  \cos( \theta_i-\theta_j)} \right]^{\half}
\ee 
where now the angles run $0 \leq \theta_k \leq \pi$ and $\tb = 2 \omega \beta = \omega \beta_c$. \\
One can also express the part of the integrand of the canonical partition function that is not coming from the measure as the determinant of a differential operator $Q$,
\be \label{IQ}
I=\left(\frac{2\pi}{\o}\right)^{\frac{1}{2}(N-2n)^2}(\det Q)^{-\frac{1}{4}}\, ,
\ee
where $Q$ is a differential operator on a circle of length $2\beta$.
\be
Q=- D_{0}^2+{\o}^2=-{\partial}_{0}^2 + 2 i\alpha{\partial}_{0}+{\alpha}^2+{\omega}^2\, ,
\ee
where $\alpha$ is a constant gauge field in the adjoint representation related to $\theta$ as $\theta _{i}=\alpha_{i}\beta$.
$Q$ acts on the matrices $M$ as
\be 
[Q,M]={\partial}_{0}M+i[\alpha ,M]
\ee
and
\be 
[\alpha ,M]_{ij}={\alpha}_{ij,kl}^{adj}M_{kl}={\alpha}_{ik}M_{kj}-M_{ik}{\alpha}_{kj}
\ee
\be 
(UMU^{\dagger})_{ij}=\exp[i\beta\alpha]^{adj}_{ij,kl}M_{kl}\, ,
\ee
with
\begin{eqnarray}
{\alpha}_{ij,kl}^{adj}={\alpha}_{ik}{\delta}_{jl}-{\alpha}_{lj}{\delta}_{ik}\, ,\qquad \exp[i\beta\alpha]^{adj}_{ij,kl}=U_{ik}U^{\dagger}_{lj}\, .
\end{eqnarray}
Thus in the momentum representation one can write
\begin{eqnarray}
\det\left(-D_{0}^2+{\o}^{2}\right)&=& \det_{matrix}\prod_{n=-\infty}^{\infty}\left[\left(\frac{2\pi n}{\beta}+{\alpha}\right)^2+{\o}^2\right]=\\
&=&\det_{matrix}\left(\cosh(\beta\o)-\cos(\beta\a)\right)\,,
\end{eqnarray}
where $\alpha$ is a matrix  and the determinant is with respect of this matrix structure.
If the gauge field is $A_{N\times N}=diag(\a_{1},\a_{2},...,\a_{n}, -\a_{1},-\a_{2},...,-\a_{n},0,...,0)$, then ${\alpha}_{ij,kl}^{adj}=({\alpha}_{i}-{\alpha}_{j}){\delta}_{ik}{\delta}_{jl}$ and \ref{IQ} equals \ref{I}.

\subsection{Measure}
One needs also to compute the measure for $\mathcal{D}U$. This is achieved by defining the metric on the tangent space of the group 
$ds^2= \mathrm{tr}(U dU^\dagger U dU^\dagger)$ and then computing its determinant to get ($ 0\leq \theta_i \leq \pi$)
\be
J_n(\theta)=\frac{1}{2^n n! (2 \pi)^n} \prod_{i<j}^n \sin^2 \left(\frac{\theta_i - \theta_j}{2} \right) \sin^2 \left(\frac{\theta_i +\theta_j}{2} \right) \prod_{k=1}^n \sin \theta_k \sin^{2(N-2n)}\left(\frac{\theta_k}{2} \right).
\ee
One finds that this is exactly the measure on the symmetric space of positive curvature defined as the coset $\frac{SU(N_1+N_2)}{SU(N_1)\times SU(N_2) \times U(1)}$ (Cartan Class $AIII$)~\cite{Caselle:2003qa} with  $N_1\equiv n  ,N_2\equiv (N-n)$. 
Again we see that $n= N/2$ is special and the measure simplifies. The normalization factor $(2\pi)^{n}$ corresponds to the stability group $U(1)^{\otimes n}$ and the factor $(2^n n!)$ to the discrete Weyl-group~\cite{Osbornnotes}.

\subsection{Pfaffian in regular representation}\label{Pfangles}

In the case of $n= N/2$ we find
\be
\mathcal{Z}_n= \int_0^\pi \prod_i d \theta_i J_n(\theta) \prod_{i , j}^n \left( \frac{4}{(\cosh \tb - \cos(\theta_i+\theta_j) (\cosh \tb - \cos( \theta_i-\theta_j)} \right)^{\half}
\ee
where the angles are in $0 \leq \theta_i \leq \pi$.

One then unfolds the denominator using for example
\be
\frac{1}{\cosh \tb - \cos(\theta_i+\theta_j)} = \frac{2 q}{(1- q  z_i z_j)(1-q z^*_i z^*_j)}, \quad q=e^{- \tb},\  z_i = e^{i \theta_i}
\ee
and similarly the measure
\be
J_n=\frac{1}{i^{n}  2^{2n^2} n! (2 \pi)^n} \prod_{i<j}^n (z_i-z_j)(z_i-z^*_j)(z^*_i-z_j)(z^*_i - z^*_j) \prod_k^n (z_k-z^*_k) 
\ee
We then define $z_i= e^{i \theta_i}$, $\bar{z}_{1..2n} = (z_{1..n}, z^*_{1..n})$ 
The partition function now is
\begin{equation}
\mathcal{Z}_n= \frac{1}{n !} \int_0^\pi \prod_{k=1}^{n}  \frac{d \theta_k}{2 \pi i}  \frac{2^{-\half}}{\sqrt{(\cosh(\tb) - \cos(2 \theta_k))}}\prod_{i<j}^{2n} \frac{q^{1/2}( \bar{z}_i-\bar{z}_j )}{1-q \bar{z}_i \bar{z}_j}
\end{equation}
From this form, one can use Schur's Pfaffian identity~\cite{Schur,Ishikawa}
\begin{equation}
 \prod_{i<j}^{2n} \frac{x_i-x_j}{1- x_i x_j} = \pf\left( \frac{x_i - x_j}{1- x_i x_j} \right)_{1 \leq i, j \leq 2n}
\end{equation}
for $x_i = q^\half \bar{z}_i$ to compactly write
\be
\mathcal{Z}_n= \frac{1}{n !} \int_0^\pi \prod_{k=1}^n \frac{ d \theta_k }{2 \pi i}\prod_{k=1}^n  \frac{  q^\half}{\sqrt{\left(1-q z_k^2 \right)\left(1-q {z^*_k}^2\right)}} \pf \begin{pmatrix}\frac{q^{1/2} (z_i- z_j)}{1-q z_i z_j}&\frac{q^{1/2} (z_i- z^*_j)}{1-q z_i z^*_j}\\\frac{q^{1/2} (z^*_i- z_j)}{1-q z^*_i z_j}&\frac{q^{1/2} (z^*_i- z^*_j)}{1-q z^*_i z^*_j}\end{pmatrix}
\ee
This is the expression that we use in the main text. This structure has appeared in connection with Ginibre's orthogonal ensemble, for more details see~\cite{Akemann:2007wa,Borodin:2007} and references within.

\section{Grand Canonical for $n=0$}
\label{n0grand}

The grand canonical partition function for $n=0$ is a partial-theta
\be
\mathcal{Z}_G = \sum_{N=0}^\infty x^N Q^{\frac{N^2}{2}}
\ee
with $Q= Z_{1}^{op}$ the 1-particle partition function with open boundary conditions and $x=e^{\beta \mu}$ the chemical potential.
Little is known about partial theta functions as compared to the usual theta functions. In~ \cite{Andrews:2005} one is able to find the proof for the following formula originally found by Ramanujan 
\bea
\label{ramanujan}
\sum_{N=0}^\infty x^N Q^{\frac{N^2}{2}} = \prod_{n=1}^\infty \left(1 - \frac{x}{f(n)} \right)  \nn \\  \frac{1}{f(n)}= -  Q^{n- \half} \left(1+ y_1(n) + y_2(n) + \mathcal{O}(Q^{\frac{3}{2}n(n+1)}) \right)
\eea
with y's computable in a recursive fashion
\bea
y_1(n)= \frac{\sum_{j=n}^\infty (-1)^j Q^{\half j(j+1)}}{\sum_{j=0}^\infty (-1)^j (2j+1) Q^{\half j(j+1)}} \nn \\
y_2(n)= \frac{\left(\sum_{j=n}^\infty (j+1) (-1)^j Q^{\half j(j+1)}\right) \left( \sum_{j=n}^\infty (-1)^j Q^{\half j(j+1)} \right)}{\left(\sum_{j=0}^\infty (-1)^j (2j+1) Q^{\half j(j+1)} \right)^2} \nn \\
y_3(n)= ...
\eea
It is amusing to note that these terms resemble the rotational partition function of diatomic molecules.
It is also easy to see that for large segment as $\beta \rightarrow \infty$, $Q \rightarrow q_c^\half$ and $y_n \rightarrow 0$ leaving
\be
\mathcal{Z}_G \approx  \prod_{n=0}^\infty \left(1+ x q_c^{\half(n+ \half)} \right)
\ee
From this expression we can correctly reproduce that the leading contribution to the free energy is  half the one of the circle in the large radius limit.
Finally, it would be interesting to study further the thermodynamic properties of equation~\ref{ramanujan}, since it is in a form (entire function) that the Lee-Yang theorem can apply. In particular a sum of positive terms does not allow for a phase transition - no zeros for x on the positive real axis, thus a phase transition is only possible if $f(n)$ can change sign for some value of $\beta$.

\section{Hilbert transform properties}\label{hilbertproperties}

In this Appendix we collect some of the properties of the Hilbert transform which can be found in~\cite{King:2009}. \\
The Hilbert transform on the real line $x\in \mathbb{R}$ of a function $f(x)$ is defined as
\be
\mathcal{H}[f](x)=\frac{1}{\pi} \mathcal{P} \int_{-\infty}^\infty \frac{f(y)dy}{x-y}
\ee
with $\mathcal{P}$ denoting the principal value.
Some properties of the transform are
\begin{itemize}
\item The Hilbert transform commutes with complex conjugation $(\mathcal{H}[f])^*= \mathcal{H}[f^*]$
\item It satisfies linearity $\mathcal{H}[a f_1 + b f_2]= a \mathcal{H}[f_1]+b \mathcal{H}[f_2]$
\item The linearity of the Hilbert transform also means that if one has a series expansion of a function $f= \sum_k f_k$ then $\mathcal{H}[f]= \sum_k \mathcal{H}[f]_k$.
\item It has the parity property of exchanging even with odd functions
\item The Hilbert transform relates the real and imaginary part of a function (Kramers-Kronig relations). As an example if $f(z)=g+i h$ is analytic in the upper half complex plane then $h(x)=-\mathcal{H}[g](x)$ and thus $\int_{-\infty}^\infty g \mathcal{H}[g] dx = 0$. Moreover $\mathcal{H}[g](x)=  h[x]$.
\item The combination with fourier transform $\mathcal{F}$ gives $\mathcal{F} \circ \mathcal{H}[f](x)=-i \sgn(x) \mathcal{F}[f](x) $
\item $\mathcal{H}^2=-I$ and thus the inverse is $\mathcal{H}^{-1}= -\mathcal{H}$.  The eigenvalues of the Hilbert transform are $\lambda = \pm i$.
\item The Hilbert transform is skew adjoint $\mathcal{H}^\dagger = - \mathcal{H}$
\item If $g(x)=\mathcal{H}[f](x)$ then $\mathcal{H}[f](ax+b)= \sgn(a)g(ax+b)$. Generically the Hilbert transform commutes with translation and positive dilations but anticommutes with reflection.
\item The Hilbert transform commutes with the derivative operator
\item The Hilbert transform commutes with $SL(2,\mathbb{R})$ generators i.e with unitary operators $U_g$ on the space $L^2(\mathbb{R})$ acting as
\be
U_{g}^{-1}f(x)=(cx+d)^{-1}f\left({ax+b \over cx+d}\right),\,\,\, \left\{ g= \begin{pmatrix}a & b \\c & d\end{pmatrix}  : a,b,c,d \in \mathbb{R}, \, ad-bc=1\right\}.
\ee
\end{itemize}

Moreover the following properties hold: \\
For an integer $n\geq 0$, $g(x)= \mathcal{H}[f](x)$
\be
\mathcal{H}[x^n f(x)]=x^n g(x)-\frac{1}{\pi}\sum_{k=0}^{n-1} x^k \int_{-\infty}^\infty t^{n-1-k} f(t) dt
\ee
Hardy:
\be
\int_{-\infty}^\infty dx \mathcal{H}[f](x) g(x) =- \int_{-\infty}^\infty dx f(x) \mathcal{H}[g](x)
\ee
Hardy-Poincare-Bertrand:
\be
\frac{1}{\pi} \mathcal{P} \int_{-\infty}^\infty \frac{f(x)dx}{x-t}\frac{1}{\pi} \mathcal{P} \int_{-\infty}^\infty \frac{g(y)dy}{y-x} =\frac{1}{\pi} \mathcal{P} \int_{-\infty}^\infty g(y)dy \frac{1}{\pi} \mathcal{P} \int_{-\infty}^\infty \frac{f(x)dx}{(x-t)(y-x)} -f(t) g(t)
\ee
One can define projection operators as follows:
\be
P_\pm = \frac{1}{2} \left(I \pm i \mathcal{H} \right)
\ee
Then one can easily see that they satisfy the properties of projection operators (idempotent conditions) $P_{\pm}^2 = P_{\pm}$.

\section{The Kernel}

\subsection{Kernel in Energy basis}\label{energykernel}

One can write down the form of the kernel in energy eigen-states and try to diagonalise from there. One has (after symmetrising appropriately):
\be
\langle m| e^{-\frac{\beta}{2} \hat H}  \mathcal{\hat O} e^{-\frac{\beta}{2} \hat H} |n \rangle=   \frac{2^{3+\frac{m+n}{2}}}{\sqrt{m! n!}} \frac{ e^{-\frac{\omega \beta}{2} (m+n+1)} \sqrt{\pi}}{n-m} \left[ \frac{1}{\Gamma(-m/2)\Gamma(\frac{-n+1}{2})} + \frac{1}{\Gamma(-n/2)\Gamma(\frac{-m+1}{2})}   \right].
\ee
To prove this formula one first has to compute $\langle m|  \mathcal{\hat O}  |n \rangle$ and it is easier to do so in momentum basis where the Hilbert transform just becomes a signum function, see appendix~\ref{hilbertproperties}
\be
\langle m|  \mathcal{\hat O}  |n \rangle = -i \int_{-\infty}^\infty dp \sgn(p) \psi_m(p) \psi_n(p) 
\ee
with $\psi_m(p)$ the Hermite functions. Note that this is non-zero only if m,n are odd/even or even/odd respectively. One can also form the diagonal components of full kernel by computing the element:
\bea\label{energykernelfull}
\langle n_1| e^{-\frac{\beta}{2} \hat H}  \mathcal{\hat O}  e^{-\beta \hat H}  \mathcal{\hat O}  e^{-\frac{\beta}{2} \hat H} |n_2 \rangle = \frac{\pi 2^{6+\frac{n_2+n_1}{2}}}{\sqrt{n_1! n_2!}}e^{-\frac{\omega \beta}{2} (n_1+n_2+1)} \times \nn \\ 
\sum_{m} \frac{ 2^{m}e^{-\omega \beta (m+1/2)}}{m!(n_1-m)(m-n_2)}  \left( \frac{1}{\Gamma(-n_1/2)\Gamma(\frac{-m+1}{2})} \frac{1}{\Gamma(-m/2)\Gamma(\frac{-n_2+1}{2})} + perm \right)\,. \nn \\
\eea
Now this kernel can be non-zero only if both $n_{1,2}$ are even or odd and the states that run through the sum are then only odd or even respectively. In either case, only one term contributes in the sum and in particular for $n_{1,2}$ odd we get ($q=e^{- \omega \beta_c}$):
\bea
 \langle n_1|\hat{\rho} |n_2 \rangle =\frac{ q^{\frac{1}{4} (n_1+n_2+2)}  2^{6+\frac{n_2+n_1}{2}}}{\Gamma(\frac{-n_1}{2}) \Gamma(\frac{-n_2}{2})\sqrt{n_1! n_2!}} \frac{{n_2}\,
   _2F_1\left(\frac{1}{2},-\frac{n_1}{2};1-\frac{n_1}{2}; q \right
   )-n_1 \,
   _2F_1\left(\frac{1}{2},-\frac{{n_2}}{2};1-\frac{{n_2}}{2}; q \right
   )}{  {n_1}^2 {n_2}- {n_1} {n_2}^2}, \nn \\
\eea
while for $n_{1,2}$ even 
\bea 
\langle n_1|\hat{\rho} |n_2 \rangle =\frac{ q^{\frac{1}{4} (n_1+n_2+2)}  2^{6+\frac{n_2+n_1}{2}}}{\Gamma(\frac{-n_1+1}{2}) \Gamma(\frac{-n_2+1}{2})\sqrt{n_1! n_2!}}
\frac{q^{\frac{{n_2}}{2}}
   B_{q}\left(\frac{1}{2}-\frac{{n_2}}{2},-\frac{1}{2}\right)-q^{\frac{{n_1}}{2}}
   B_{q}\left(\frac{1}{2}-\frac{{n_1}}{2},-\frac{1}{2}\right)}{4  
   ({n_1}-{n_2})}. \nn \\
\eea
which can also be rewritten in terms of $ _2F_1$. From this expression we can also match the formulas in~\ref{lbregular} for $\mathcal{\hat O}^2$ if we set $\beta=0$.

\subsection{Kernel in elliptic functions}\label{elliptickernel}
One can massage a bit the integral equation~\ref{elliptickernelfinal}, by adding/subtracting information from both sheets. In terms of the torus this means to form (the parentheses in both sides of the equation stand for the even/odd case)
\be
\lambda\left(  X(u) \left( \pm \right) X(u+2K) \right)= - 2 q^\half \int_{C_1+C_2} \frac{ dv}{2 \pi i}  \begin{pmatrix}
q \sn v \cn^2 u \\ \sn u \dn^2 v
\end{pmatrix} \frac{X(v)}{\dn^2 v - \cn^2 u},
\ee
where the denominator can be also written as $\sn^2 u - q^2 \sn^2 v$. One can bring this equation into the following final form
\be
\lambda  X_{(\pm)}(u) = -  q^\half \int_{C_1+C_2} \frac{ dv}{2 \pi i}  \begin{pmatrix}
q \sn v \cn^2 u \\ \sn u \dn^2 v
\end{pmatrix} \frac{X_{(\pm)}(v)}{\dn^2 v - \cn^2 u}
\ee
with $ X_{(\pm)}(u)=X(u) \left( \pm \right) X(u+2K)$.

\subsection{Trace of the kernel}
\label{trkernel}

The trace of the kernel can be computed to be (also using equation~\ref{rho})
\begin{equation}
\tr \hat \rho =\frac{1}{ \sqrt{2} \sinh(\tb /2)} \int_0^\pi \frac{ d \theta}{2 \pi} \frac{\sin{\theta}}{\sqrt{\cosh \tb - \cos(2\theta)}} = \frac{1}{2 \pi \sinh(\tb /2)} \tan^{-1} \frac{1}{\sinh(\tb /2)}
\end{equation}
Due to the branch-cut structure of this expression, it is useful to represent this function in terms of an integral with the integrand having simple poles
\begin{equation}
\tr \hat{\rho}= \frac{1}{2 \pi \sinh (\tb/2)} \arctan \left( \frac{1}{\sinh (\tb/2)} \right)= \int_{0}^{\pi} \frac{d \theta}{2 \pi} \frac{\cos(\theta/2)}{\cosh(\tb)- \cos(\theta)}
\end{equation}
(keep in mind that $\tb= \omega \beta_{circle} = 2 \omega \beta_{orb}$). To discuss the inverse oscillator one needs to set $\omega \rightarrow - i \omega$. One then finds
\begin{equation}
\tr \hat{\rho}_{inv}= \frac{- 1}{2 \pi \sin (\omega \beta_c /2)} \tanh^{-1} \left( \frac{1}{\sin (\omega \beta_c /2)} \right)= \int_{0}^{\pi} \frac{d \theta}{2 \pi} \frac{\cos(\theta/2)}{\cos(\omega \beta_c)- \cos(\theta)}
\end{equation}
An analogous formula for the circle is~\cite{Boulatov:1991xz}
\begin{equation}\label{anglecirclepf}
Z_{circ}^{inv}(\beta_c) =\sum_{k=0}^\infty e^{i \omega \beta_c (k+\half)} = \frac{i}{2 \sin (\omega \beta_c /2)} = \int_{0}^{\pi} \frac{d \theta}{2 \pi} \frac{1}{\cos(\omega \beta_c/2)- \cos(\theta)}
\end{equation}
If we define the twisted partition function~\cite{Boulatov:1991xz}
\begin{equation}
Z(\theta, \beta_c ) = \frac{1/2}{\cos \omega \beta_c - \cos \theta}
\end{equation}
we understand both results as a 1-particle partition function derived from  averaging over twist angles with a different weight for the orbifold and circle (after extending due to symmetry the integrals for $\theta' \in [-\pi, \pi]$).
Another useful representation is
\begin{equation}
Z(\theta,  \beta_c ) = \int ^{+\infty} _{-\infty } d\epsilon \ {\rm e}^{-\beta_c
 \epsilon }\rho (\theta, \epsilon ) = {1 \over \sin \theta} \int ^{+\infty} _{-\infty }  d\epsilon \
{\rm e}^{-\beta_c \epsilon } {\sinh {\frac{\epsilon}{\omega}  }(\pi  - \theta) \over 
\sinh {\frac{\epsilon}{\omega}   } \pi }
\end{equation}
which holds for
$0 < \theta < 2\pi$ and $\rho(\theta, \epsilon)$ is the twisted density of states.
From this one finds a closed formula for the twisted dos:
\begin{equation}
\rho (\theta, \epsilon ) = {\sinh { \frac{\epsilon}{\omega} }(\pi  - \theta) \over
\sinh { \frac{\epsilon}{\omega}  }\pi
\sin \theta}
\end{equation}
and also an expression that gives away the spectrum
\begin{equation}
\rho (\theta,\epsilon ) = \sum^\infty _{m=-\infty }e^{im\theta}\rho ^{(m)}(\epsilon)
  = {1 \over \pi } \sum^\infty _{k=0} \sum^\infty _{m=-\infty }
{e^{im\theta}({\mid m \mid + 1 \over 2} + k ) \over (\frac{\epsilon}{\omega})^2 +
(k+ {\mid m \mid + 1 \over 2})^2} +
\delta (\theta)\log \Lambda ^2
\end{equation}
note in particular the logarithmic divergence at $\theta =0$ that is regulated putting a wall at some cutoff $\Lambda$ and neglecting any cutoff dependent quantities in the double scaling limit. In this equation $\rho^m(\e)= -\frac{1}{\pi} Re \Psi(i\frac{\e}{\omega} +\frac{|m|+1}{2})$ is the Hydrogen atom density of states (discrete spectrum) which should be contrasted with the H.O. density of states $\rho_{H.O.}(\e)=- \frac{1}{2\pi} Re \Psi(i \frac{\e}{\omega} +\half)$.

\subsection{1-particle density of states}
From the partition function $Z(\beta)$, one computes the density of states using
\begin{equation}
\rho_d(\epsilon) =\int_{c -i \infty}^{c + i \infty} \frac{d \beta}{2 \pi i} Z(\beta) e^{\beta \epsilon}
\end{equation}
The difficulty in our case is that one needs again to study very well the pole and branch cut structure of the integrand. We will instead try to use the integral representation for the partition function of the orbifold to write
\begin{equation}
\rho_o(\epsilon) = \frac{1}{2} \int_{c -i \infty}^{c + i \infty} \frac{d \beta_c}{2 \pi i }  \int_{0}^{\pi} \frac{d \theta}{2 \pi} \frac{\cos(\theta/2)}{\cos(\omega \beta_c)- \cos(\theta)} e^{\beta_c \epsilon}
\end{equation}
with c an infinitesimal positive regulator. Interchanging the integrations one picks the poles at the negative $\beta_c$ axis $\beta_c= 2 n \pi \pm \theta$ and sums over the residues. There is a catch when $\theta \rightarrow 0$, since then two poles merge and the singularity pinches the contour. In any case, the same singularity appears also in the analogous formula of the circle~\ref{anglecirclepf} and will just reproduce the irrelevant logarithmic divergence. The result is
\begin{equation}
\label{dos}
\rho_o(\epsilon)  = \int_{0}^{\pi} \frac{d \theta}{2 \pi} \frac{\cos(\theta/2)}{\sin(\theta)}\frac{\sinh \frac{\epsilon}{\omega} (\pi - \theta)}{\sinh \frac{\epsilon}{\omega} \pi} =   \int_0^\pi \frac{d \theta}{2 \pi} \cos(\theta/2) \rho(\theta , \epsilon)
\end{equation}
It is thus easy to see that this result is equivalent to the one we would get if we just integrate over the twisted dos with the appropriate weight. Now this integral can be performed indefinite to get a result in terms of hypergeometric functions $_2F_1$. Taking the limit $\theta \rightarrow 0$ and subtracting the expected logarithmic divergence, we find a finite piece
\bea
\rho^0 (\e) & = & \frac{1}{4 \pi} \left( i \pi - 2\gamma  + \frac{e^{- \pi \frac{\epsilon}{\omega}}}{\sinh( \pi \frac{\epsilon}{\omega})} \Psi \left(-i
  \frac{\epsilon}{\omega} +\frac{1}{2} \right)- \frac{e^{ \pi  \frac{\epsilon}{\omega}}}{\sinh( \pi \frac{\epsilon}{\omega})}\Psi \left( i \frac{\epsilon}{\omega} +\half \right)\right)  \nn \\
& = & \frac{i}{2} - \frac{1}{2 \pi} \gamma  -  \frac{1}{2 \pi} Re \Psi \left(
   i \frac{\epsilon}{\omega} +\frac{1}{2}\right) + i  \frac{1}{2 \pi} \frac{\cosh( \pi \frac{\epsilon}{\omega})}{\sinh( \pi \frac{\epsilon}{\omega})} Im \ \Psi \left( i \frac{\epsilon}{\omega} +\half\right)
\eea
that contains the H.O. dos $\rho_{HO}(\epsilon)=-  \frac{1}{2 \pi} Re \Psi \left(
   i \frac{\epsilon}{\omega} +\frac{1}{2}\right) $, and imaginary pieces.
From the $\pi$ limit we get\footnote{One nice thing to note is that the twisted part of the dos does not require a cutoff in accordance with the discussion in~\cite{Boulatov:1991xz}.}
\begin{equation}
\rho^{\pi}(\epsilon) = \frac{1}{4 \pi \sinh(\frac{\epsilon}{\omega} \pi)} \left[ Im \ \Psi \left( i \frac{\epsilon}{2 \omega} +\frac{1}{4})\right) - Im \ \Psi \left( i \frac{\epsilon}{2\omega} +\frac{3}{4})\right)\right]
\end{equation}
One then notices that the 1-particle orbifold density of states is $\rho_o(\epsilon)= \rho_{H.O.}(\epsilon) + \rho_{twisted} +  \rho_{Im}(\e)$, with the twisted piece
\bea\label{twisteddos}
\rho_{twisted}(\epsilon) = \frac{1}{4 \pi \sinh(\frac{\epsilon}{\omega} \pi)} \left[ Im \ \Psi \left( i \frac{\epsilon}{2 \omega} +\frac{1}{4})\right) - Im \ \Psi \left( i \frac{\epsilon}{2\omega} +\frac{3}{4})\right)\right] \nn \\
= \frac{1}{4 \pi \sinh(\frac{\epsilon}{\omega} \pi)} Im \ \int_0^\infty dt \frac{e^{-i \frac{\epsilon}{\omega} t}}{\cosh(\frac{t}{2})}
\eea
where we used the integral representation of the digamma function.

\section{Approximate methods for large $\beta$}\label{apapprox}
Here we give more details on the large $\beta$ approximation to the canonical partition function.


\subsection{Generic $n$ in angles}\label{lbgeneric}
One can expand eq.~\ref{ncanonical} for large $\beta$ and relating $q_c=q_o^2$ to find
\bea
\mathcal{Z}_n \approx q_c^{\frac{ (N-2n)^2}{4}} q_c^{N-2n} q_c^{n^2}(1+\mathcal{O}(q_c))  \int_0^\pi \prod_k d \theta_k  J_n(\theta) \nn \\
= q_c^{\frac{N^2}{4}} (1+\mathcal{O}(q_c)) \frac{1}{n!} \prod_{j=0}^{n-1} \frac{\Gamma(1+j) \Gamma(2+j) \Gamma(N-2n+1+j)}{\Gamma(N-n+j+1)}
\eea
where we  used again the Selberg integral to compute the integral. We find that the leading in $\beta_c$ term will give half the Free-energy of the circle for any n.

\subsection{Generic $n$ in  eigenvalues of $M$}
\lab{genericnlambda} 

We start by the following generalization of the Cauchy identity~\cite{Basor}
\begin{align}
\frac{\prod_{i<j}^{n} (x_i-x_j) \cdot
\prod_{a<b}^{N-n} (y_{a}-y_{b})}
{\prod_{i=1}^{n} \prod_{a=1}^{N-n} (x_i -y_{a})} 
= (-1)^{n (N-2n)}
\det \begin{pmatrix}
\tfrac{1}{x_1-y_1} & \cdots & \tfrac{1}{x_1-y_{N-n}} \\
\vdots & \ddots & \vdots \\
\tfrac{1}{x_{n}-y_1} & \cdots & \tfrac{1}{x_{n}-y_{N-n}} \\
y_1^{N-2n-1} & \cdots & y_{N-n}^{N-2n-1} \\
\vdots & \ddots & \vdots \\
y_1^0 & \dots & y_{N-n}^0 
\end{pmatrix}.
\label{eq:det}
\end{align}
where on the right hand side, the upper $N\times N-n$ submatrix and
the lower $(N-2n)\times n$ submatrix are given respectively by
\begin{align}
\biggl(\frac{1}{x_i-y_a}\biggr)
_{\begin{subarray}{c} 1\leq i\leq n\\1\leq a\leq N-n\end{subarray}},\quad
\bigl(y_a^{N-2n-p}\bigr)
_{\begin{subarray}{c} 1\leq p\leq N-2n\\1\leq a\leq N-n\end{subarray}}\, .
\end{align}
One can now perform the y integrations to obtain
\bea
\int d^{n} x d^{N-n} y \det_{N-n \times N-n} \begin{pmatrix} \biggl(\frac{1}{x_i-y_a}\biggr)
_{\begin{subarray}{c} 1\leq i\leq n\\1\leq a\leq N-n\end{subarray}} \\
\bigl(y_a^{N-2n-p}\bigr)
_{\begin{subarray}{c} 1\leq p\leq N-2n\\1\leq a\leq N-n\end{subarray}}
\end{pmatrix}  \det_{N \times N} \psi_{i-1}(\bar{x}_j) \, .
\eea
It is reassuring to check that the formula reproduces correctly the cases of $n=0$ and $n=N/2$. One can then perform one extra integration to reach the formula eqn.~\ref{generictheta} of the main text.

\subsection{n=0}\label{lbn0}

For the $n=0$ representation, we define $Dx = \prod_k^N dx_k/N!$ and expand in multi-particle fermionic wavefunctions
\bea
Z_N= \int Dx Dy \Delta(x) \Delta(y) \det_{i j} K(x_i , y_j)=  \int Dx Dy \Delta(x) \Delta(y) \prod_k K(x_k , y_k) = \nn \\
= \int Dx Dy \Delta(x) \Delta(y) \sum_{E_{n}} e^{- \beta E_{n}} \Psi_{E_{n}}(x_k) \Psi_{E_{n}} (y_k)
\eea
with $\Psi_{E_{n}} (y_k)$ the multiparticle energy eigenfunctions $\langle E_{n} | y_1, y_2, ....y_N \rangle$.
This in the $\beta \rightarrow \infty$ limit gives
\bea
Z=e^{- \beta E_{ground}} \left( \int Dx \Delta(x)  \Psi_{ground}(x_k) \right)^2 &=&e^{- \beta E_{ground}} \left( \int Dx \det_{ i, k}( x_k^{i-1} ) \det_{j, k} (  \psi_{j-1}(x_k) ) \right)^2  \nn \\
&=& e^{- \beta E_{ground}} \left(\det_{i j} \int dx ( x^{i-1} \psi_{j-1}(x) ) \right)^2 
\eea
with $\psi_i(x_k)$ the single-particle wavefunctions and we used Andreief identity~\cite{deBruijn:1955} to turn the integral over N variables to an integral over a single one.
The Free energy is
\bea
\mathcal{F}= + \half \beta_c E_{ground} - 2 \log \det_{0 \leq i,j \leq N} \int dx ( x^{i-1} \psi_{j-1}(x) )
\eea
where the second term can be interpreted as a radius independent contribution of states at the endpoints written as a determinant of a matrix $F_{i  j}$. One needs to compute the following integrals
\bea
F^+_{n m}= \int_{-\infty}^\infty dx x^{2n-2} \psi_{2m-2}(x)  \rightarrow \int_{-\infty}^\infty dx x^{2n-2} \psi^+(\e_{m-1}, x)  \nn \\
F^-_{n m} = \int_{-\infty}^\infty dx x^{2n-1} \psi_{2m-1}(x) \rightarrow \int_{-\infty}^\infty dx x^{2n-1} \psi^-(\e_{m-1} , x)
\eea
where we have indicated the corresponding expressions for the normal and the inverse H.O. To compute this contribution for the inverse harmonic oscillator we will use the odd/even parabolic cylinder $\psi^\pm$ functions of appendix~\ref{parabolic}.
We define $\a =\frac{1}{4}-i\frac{\e}{2}$ and use the following integral (c is an infinitesimal regulating parameter)
\bea
I&=&\int^{\infty}_{0}dx x^{2n}e^{-i \frac{x^2}{4}}{_1F_{1}}\le(\a ;\g ;\frac{i x^2}{2}e^{i c}\ri)=\nn\\
&=& 2^{2n}e^{\frac{i\pi}{4}(2n+1)}\Gamma (n+\frac{1}{2}){_2F_{1}}\le(\a ;n+\frac{1}{2};\g ;2e^{i c}\ri)
\eea
This can be proven using Mellin-Barnes representations for hypergeometric functions. We then get
\bea
F^{+}_{n m}&=&2^{2n-1}e^{\frac{i\pi}{4}(2n-1)}C_{+}(\e)\Gamma(n-\frac{1}{2}){_2F_{1}}\le(\a , n-\frac{1}{2} ;\frac{1}{2}; 2 e^{i c}\ri) \\
F^{-}_{n m}&=&2^{2n+1}e^{\frac{i\pi}{4}(2n+1)}C_{-}(\e)\Gamma(n+\frac{1}{2}){_2F_{1}}\le(\a+\frac{1}{2} , n+\frac{1}{2} ;\frac{3}{2}; 2 e^{i c}\ri)\, ,
\eea
using the following identity for the hypergeometric functions
\be 
F\le(a,b;c;z\ri)=(1-z)^{c-a-b}F\le(c-a,c-b;c;\frac{z}{z-1}\ri)
\ee
we find
\bea
F^{+}_{mn}&=& 2^{2n-1}e^{\frac{i\pi}{4}(2n-1)}C_{+}(\e_{m-1})(-1)^{n-1}e^{i\pi\alpha}\Gamma(n-\frac{1}{2}){_2F_{1}}\le(\frac{1}{2}-\a ,1- n ;\frac{1}{2}; 2 e^{-i c}\ri)= \nn\\
&=& 2^{2n-1}e^{\frac{i\pi}{4}(2n-1)}C_{+}(\e_{m-1})(2)^{n-1}e^{i\pi\alpha}\sqrt{\pi}\nn\\
&&\times \sum_{k=0}^{n-1}\frac{\Gamma\le(\frac{1}{2}-\a+n-1-k\ri)}{\Gamma\le(\frac{1}{2}-\a\ri)}\frac{\Gamma\le(n-\frac{1}{2}\ri)}{\Gamma\le(n-\frac{1}{2}-k\ri)}\frac{\le(n-1\ri)!}{\le(n-1-k\ri)!}\frac{\le(-\frac{1}{2}\ri)^{k}}{k!}=\nn\\
&=&\frac{2^{2n-1}e^{\frac{\pi}{4}\e_{m-1}}|\Gamma\le(\a_{m-1}\ri)|}{2^{\frac{5}{4}}\sqrt{\pi}}\sum_{k=0}^{n-1}a_{k}(n)\e_{m-1}^{n-1-k}
\eea
where $a_{k}$ depends only on $n$. Similarly,
\bea
F^{-}_{n m}&=& 2^{2n+1}e^{\frac{i\pi}{4}(2n+1)}C_{-}(\e_{m-1})(-1)^{n-1}e^{i\pi(\alpha+\frac{1}{2})}\Gamma(n+\frac{1}{2}){_2F_{1}}\le(1-\a ,1- n ;\frac{3}{2}; 2 e^{-i c}\ri)= \nn\\
&=& 2^{2n+1}e^{\frac{i\pi}{4}(2n+1)}C_{-}(\e_{m-1})(2)^{n-1}e^{i\pi (\alpha+\frac{1}{2})}\frac{\sqrt{\pi}}{2}\nn\\
&&\times \sum_{k=0}^{n-1}\frac{\Gamma\le(\frac{1}{2}-\a+n-\frac{1}{2}-k\ri)}{\Gamma\le(1-\a\ri)}\frac{\Gamma\le(n+\frac{1}{2}\ri)}{\Gamma\le(n+\frac{1}{2}-k\ri)}\frac{\le(n-1\ri)!}{\le(n-1-k\ri)!}\frac{\le(-\frac{1}{2}\ri)^{k}}{k!}=\nn\\
&=&\frac{2^{2n}e^{\frac{\pi}{4}\e_{m-1}}|\Gamma\le(\a_{m-1}+\frac{1}{2}\ri)|}{2^{\frac{3}{4}}\sqrt{\pi}}\sum_{k=0}^{n-1}b_{k}(n)\e_{m-1}^{n-1-k}
\eea
with $b_{0}=1$. After using determinantal properties, we find that
\bea
\ln\le(\det F^{+}_{mn}\ri)&=&\sum_{i<j}\ln\le(\e_{i-1}-\e_{j-1}\ri)+\sum_{i}f_{i}\\
\ln\le(\det F^{-}_{mn}\ri)&=&\sum_{i<j}\ln\le(\e_{i-1}-\e_{j-1}\ri)+\sum_{i}g_{i}
\eea
with,
\bea
f_{i}= \frac{\pi}{4}\e_{i-1}+\ln\vert\Gamma\le(\frac{1}{4}-\frac{i\e_{i-1}}{2}\ri)\vert\, ,\qquad 
g_{i}= \frac{\pi}{4}\e_{i-1}+\ln\vert\Gamma\le(\frac{3}{4}-\frac{i\e_{i-1}}{2}\ri)\vert\,.
\eea
Note that as $\e\rightarrow\infty$
\be 
f(\e)+g(\e)=\ln (2\pi)-\frac{1}{2}\ln\le(1+e^{-2\pi\e}\ri)=\ln(2\pi)-\frac{1}{2}e^{-2\pi\e}+...
\ee
contributing only non perturbative terms.\\
Introducing the density of states, one obtains a quite simple result for the twisted state contribution
\be\label{tw} 
\Theta = \frac{1}{2}\int^{\mu}\rho (\e)\int^{\mu}\rho (\e^{\prime})\log\vert\e -\e^{\prime}\vert d\e d\e^{\prime} 
\ee
where $\rho (\e)$ is the density of states:
\be 
\rho (\e)=\frac{1}{\pi}\le(-\log\e +\sum_{m=1}^{\infty}\mathcal{C}_{m}\e^{-2m}\ri)
\ee
the coefficients $\mathcal{C}_{m}$ are known in terms of Bernoulli numbers. 
To compute this quantity we take one derivative wtr to $\mu$ to get
\be
\frac{\partial \Theta}{\partial \mu} = \rho(\mu) \int_{- \infty}^0 d \e \rho(\e + \mu) \log|\e| 
\ee \\
In this expression one needs to put a cutoff $\Lambda$ at the lower part of integration and compute it as a series expansion in $1/\mu$.
After one computes \ref{tw}, one has to express it in terms of the cosmological constant $\Delta$ in order to be able to compare with the Liouville result (see section~\ref{lsegment}). One needs to use
\be 
\frac{\partial\Delta}{\partial\mu}=\pi\rho(\mu)\, ,
\ee
and the renormalised cosmological constant $\mu_{0}$ that plays the role of the string coupling,  defined via
\be  
\Delta=- \mu_{0}\log\mu_{0}\, .
\ee
In the end $\Theta$ can be found in terms of $\mu_{0}$ as:
\bea
\Theta &=&\mu_{0}^{2}\le(\frac{11}{8}-\frac{\pi^{2}}{24}+\le(\frac{\pi^{2}}{12}-\frac{11}{4}\ri)\log{\mu_{0}}+\frac{7}{4}\log^{2}\mu_{0}-\frac{1}{2}\log^{3}\mu_{0}\ri)\nn\\
&&-\frac{1}{24}\le(1+\frac{\pi^{2}}{6}\ri)\log\mu_{0}+\frac{1}{\mu_{0}^{2}}\le(\frac{259}{11520}+\frac{7}{2880}\le(\frac{\pi^{2}}{3}-7\ri)\log\mu_{0}\ri)\mathcal{O}(\mu_{0}^{-4})\,. \nn \\
\eea
One notices that the torus contribution is not the same as in equation~\ref{orblfinal}.

\subsection{$n= N/2$ with Hermite polynomials}\label{lbregular}

For the regular case we get (the measures contain appropriate factorials)
\bea
Z= \int dx dx' dy dy' \det_{i, j}\frac{1}{x_i- y_j} \det_{i, j}\frac{1}{x'_i- y'_j} \det_{2n \times 2n} K(x,y ; x',y') 
\eea
which in the $\beta \rightarrow \infty$ limit gives
\bea
\mathcal{F}=\half \beta_c E_{ground}  - 2 \log \int d^n x  d^n y \det_{i, j}\frac{1}{x_i- x'_j}  \det_{2n \times 2n} \psi_{i-1}(\bar{x}_j) 
\eea
with $\bar{x}= (x,y)$. One can use Moriyama's formula for unequal ranks in the appendix of~\cite{Matsumoto:2013nya} to get
\be
\Theta= 2 \log \int d^n x \det_{\begin{subarray}{c} 1 \leq i \leq 2n \\ 1 \leq k \leq n \end{subarray}} \left[\int dy \frac{\psi_{i-1}(x_k)}{x_k-y} \quad \psi_{i-1}(x_k) \right]
\ee
As we have dicussed, one can also integrate x's to find
\be
\Theta= 2 \log \left[ \pf_{2n \times 2n} O_{i j} \right]
\ee
with the antisymmetric
\bea
O_{i j}= \int dx dy \frac{\psi_{i-1}(x)\psi_{j-1}(y)-\psi_{i-1}(y)\psi_{j-1}(x)}{x-y} =2 \int dx dy \frac{\psi_{i-1}(x)\psi_{j-1}(y)}{x-y}
\eea
 Similarly to the main text we will adopt the principal value prescription. This gives
\be
O_{i, j}= \int_{-\infty}^\infty dx \psi_{i-1}(x) \psi^\mathcal{H}_{j-1}(x) -\psi^\mathcal{H}_{i-1}(x) \psi_{j-1}(x) = 2  \int_{-\infty}^\infty dx \psi_{i-1}(x) \psi^\mathcal{H}_{j-1}(x) 
\ee
with $\psi^\mathcal{H}$ the Hilbert transform of $\psi$ and in the second line we used that the Hilbert transform is skew-adjoint. For the Pfaffian  we have the formula $\log\left[ \pf A \pf B \right]= \half \tr \log A^T B$. where we want to apply it for the case $A=B=O$ with $O^T= -O$  so that we get
\be
\Theta= \half \tr \log \left(  -O^2 \right)
\ee
One notices that the matrix O is just twice the Hilbert transform  operator $\mathcal{\hat{O}}$ in the energy basis. It is a real antisymmetric matrix with imaginary eigenvalues. Also, since $\mathcal{H}^2=-1$ (see appendix~\ref{hilbertproperties}), we immediately find $\Theta= \half \tr \log 4 \hat I  = N \log 2$.
To be more explicit, if we perform the integrals we can rewrite O as:	
\be
O_{m, n}=2 \langle m | \hat{\mathcal{O}} | n \rangle  =\pm 4 \frac{2^{2+\frac{m+n}{2}}}{\sqrt{m! n!}} \frac{\sqrt{\pi}}{n-m} \left[ \frac{1}{\Gamma(-m/2)\Gamma(\frac{-n+1}{2})} + \frac{1}{\Gamma(-n/2)\Gamma(\frac{-m+1}{2})}   \right]
\ee
with $0 \leq m,n \leq N-1$. In this expression, only one of the two terms inside the brackets can be non-zero when $m$-odd, $n$-even or vice versa, the odd/odd even/even pieces are zero. The overall $\pm$ is because the hermite functions are eigenfunctions of the fourier transform with eigenvalues $\pm 1, \pm i$ and one finds an overall factor $(-i)^{m+n+1}$, when going to momentum space in order to calculate the integral.   

Using this we can form $O^2$ as (this now holds for $n_1, n_2$ together odd/even!)
\be
O^2_{n_1 n_2} = \frac{2^{n_1/2+n_2/2+3} \sqrt{\pi}}{\sqrt{n1! n2!}(n_1-n_2)} \left[\frac{1}{\Gamma(-\frac{ n_1}{2})\Gamma(\frac{-n_2+1}{2})} - \frac{1}{\Gamma(- \frac{n_2}{2})\Gamma(\frac{-n_1+1}{2})}\right]
\ee
In this expression, we find that the only non-zero terms are the diagonal. This is also consistent with the appropriate limit of the full energy-basis kernel~\ref{energykernelfull}. Near the diagonal this expression approaches the \emph{sine-kernel}
\be\label{sinekernelnormal}
O^2_{n_1 n_2} \approx - \frac{4 \sin \pi (n_1 - n_2)}{\pi (n_1 - n_2)}
\ee
Taking the limit $n_2 \rightarrow n_1$ we find 
\be
\Theta= \half \tr \log \left( - O^2 \right) = \half \sum_{k=0}^{N-1} \log\left[ \left( \Psi(-k/2) - \Psi(1/2-k/2)\right)\frac{2 \sin \pi k }{\pi }\right]
\ee
The expression in brackets has only real part. One also finds
\be
\lim_{k \rightarrow \mathbb{N}} \left( \Psi(-k/2) - \Psi(1/2-k/2)\right)\frac{2 \sin \pi k }{\pi } = 4, \quad \forall k \in \mathbb{N},
\ee
and thus we recover the expected $\Theta= N \log 2$ which pinpoints to the fact that we just count the total entropy of a two state system at the endpoints, due to the spin up/down nature of the wavefunctions. It is tempting to pass to continuous variables via the dos $\rho_{H.O.}(\e)= - \frac{1}{\pi} \sum_k \delta(\e - \e_k)$  which for the inverse H.O. clicks when $-i \e = k+ \half$.
The result is
\bea
\Theta=  \half \int^\mu d \e \rho_{H.O.}(\e) \log \left[ \left( \Psi(\frac{1}{4}+  \frac{i \e}{2}) - \Psi(\frac{3}{4}+  \frac{i \e}{2} \right)\frac{ 2 \cosh (\pi \e)}{\pi} \right]  ,
\eea
with the term in the logarithm looking conspicuously similar to the twisted dos equation~\ref{twisteddos}. One should be very careful though, since the normalization of the Hermite functions after rotating is different compared to the one of the parabolic cylinder functions and one should really perform the computation from the start using the inverse H.O. eigenfunctions.

\subsection{$n= N/2$ with parabolic cylinder functions}\label{lbregularparabolic}

Here we perform the same computation using the delta-function normalised even and odd parabolic cylinder functions of appendix~\ref{parabolic} which are eigenfunctions of the inverted oscillator. Since the spectrum is now continuous, we can imagine obtaining a discrete spectrum by putting a cutoff/wall at $\Lambda$  which is then send to infinity. We again adopt the principal value prescription whenever fourier transforming.
\\
\\
We compute\footnote{Only the energy dependence is important in the overall normalisation of this object.}
\be
\langle \e_1 | O |\e_2 \rangle = 2 \int_{-\infty}^\infty dx dy \frac{\psi^+(\e_1, x)\psi^-(\e_2, y)}{x-y} = 4  \int_{0}^\infty d x \int_{0}^\infty d y \frac{ \psi^+(\e_1,x) y \psi^-(\e_2,y)}{x^2 - y^2} \, .
\ee
This expression is non zero and the integrand is even both in $x$ and $y$. One can then exponentiate again the denominator using the Fourier transform of the sign function. This gives
\be  
O(\e_1, \e_2) = -2 i \int_{-\infty}^\infty d t \sgn (t) I^+(t) I^-(t) = -4 \Re \left[ i \int_0^\infty dt I^+(t) I^-(t) \right]
\ee
where,
\be
I^+(t) = \int_0^\infty d x \psi^+(x) e^{-i \half t x^2} \, , \quad I^-(t) = \int_0^\infty d y y \psi^-(y) e^{+i \half t y^2}\,.
\ee
The advantage is that now one can compute the resulting integrals using~\cite{Jeffrey:2007}
\bea\label{hyper1}
\int_0^\infty d u e^{- s u} u^{b-1}\, _1F_1(a, c , k u) &=& \Gamma(b) s^{- b}\, _2F_1 (a, b, c , k s^{-1}),\nn\\&&\Re{s} > \Re{k} ,\, \Re{s} > 0 , \,  b>0 \, ;;|s|>|k| \,,\nn\\
&=&\Gamma(b) (s-k)^{- b}\, _2F_1 (c-a, b, c , \frac{k}{k- s}),\nn\\
&&\Re{s} > \Re{k},\,\Re{s} > 0,\,b>0 \, ;; |s-k|>|k|\nn 
\eea
Or even the following simpler form that can be obtained from the expression above if $k=1$, $b=c$
\be\label{hyper2}
\int_0^\infty d u e^{- s u} u^{c-1}\, _1F_1(a, c ,  u) = \Gamma(c)s^{-c}(1-s^{-1})^{-a} \, , \qquad  \Re{c}>0 \, , \quad  \Re{s}>1\, .
\ee
Using an infinitesimal regulator $e^{i c}$ we can find for $I^{+}(\epsilon_{1},t)$ with $t>\half$
\bea\label{Iplus}
I^+ (\e_1 , t) &=& N_1(\e_1) \int_0^\infty \frac{d u}{\sqrt{2 u}} e^{-i(\half+t) u e^{i c}} {_1F_1} (1/4-i \e_1 /2 , 1/2 ; i u e^{i c}) \nn \\
&=& N_1(\e_1) \sqrt{\frac{\pi}{2}} (it + \frac{i}{2})^{- 1/2} \, _2F_1(1/4-i \e_1 /2, 1/2 , 1/2 , \frac{e^{i c}}{\half + t}) \, ,
\eea
with $N_1(\e)= \left(\frac{1}{4 \pi \sqrt{(1+e^{2 \pi \e})}}\right)^\half 2^{1/4} \lvert \frac{\Gamma(1/4+ i \e/2)}{\Gamma(3/4+ i \e/2)} \rvert^\half$.
\\
\\
For $I^{-}(\epsilon_{2}, t)$,  we now have  (with $t< -\half$),
\bea\label{Iminus}
I^- (\e_2 , t) &=& N_2(\e_2) \int_0^\infty d u \sqrt{2 u}  e^{-i(\half-t) u e^{i c}} {_1F_1} (3/4-i \e_2 /2 , 3/2 ; i u e^{i c}) \nn \\
&=& N_2(\e_2) \sqrt{\frac{\pi}{2}} (-it + \frac{i}{2})^{- 3/2} \, _2F_1(3/4-i \e_2 /2, 3/2 , 3/2 , \frac{e^{i c}}{\half - t}) \, ,
\eea
with $N_2(\e) = \left(\frac{1}{4 \pi \sqrt{(1+e^{2 \pi \e})}}\right)^\half  2^{3/4}  \lvert \frac{\Gamma(3/4+ i \e/2)}{\Gamma(1/4+ i \e/2)} \rvert^\half$. We now encounter a form of non-perturbative ambiguity which has to do with the possible analytic continuations of these hypergeometric functions. In particular, the hypergeometric functions $_2 F_1(a,b,c,z)$ have branch points at $z=(0,1,\infty)$ and thus the integrals~\ref{Iplus},~\ref{Iminus} have branch points at $t=(\infty,\half, - \half)$ and $t=(-\infty, -\half, \half)$ respectively. We will now assume working in some undetermined  branch and naively analytically continue these equations for complex $t$. In the next subsection we are going to split the $t$ integral into sections and find what are the exact conditions (which sheet to choose) in order to match the result we find here.

We will now introduce the following change of variables $z=1/(\half+t)\, , t= (2-z)/2z$ to get
\bea
&&O(\e_1, \e_2) = - 2  \pi  N_1(\e_1) N_2(\e_2) \times \nn \\
&&\Re \left[ \int_{0}^2 \frac{d z}{z^2} \left[ \frac{z}{z-1} \right]^{\frac{3}{2}} \left[ z \right]^{\half} \,  _2F_1(3/4-i \e_2 /2, 3/2 , 3/2 , \frac{z e^{i c}}{z-1})\,  _2F_1(1/4-i \e_1 /2, 1/2 , 1/2 , z e^{i c}) \right] \, \nn \\
\eea
By shifting the corresponding hypergeometric function and performing the integral we get
\bea\label{hypergeometricresult}
O(\e_1, \e_2) & = & -2 \pi    N_1(\e_1) N_2(\e_2)  e^{\half(\e_1 + \e_2)\pi}
\Re \left[ i \int_{0}^2 d z (z-1)^{-1+ \half i (\e_1-\e_2)}  \right] \, \nn \\
&=& - 2 \pi  N_1(\e_1) N_2(\e_2) e^{\half(\e_1 + \e_2)\pi} \Re \left[ \int_{-1}^1 \frac{du}{u} u^{+ \half i (\e_1-\e_2) }   \right]\,  \nn \\
& = & 4 \pi  N_1(\e_1) N_2(\e_2) e^{\half(\e_1 + \e_2)\pi} \left[ \frac{1-e^{\half(\e_2 - \e_1)\pi}}{\e_1 - \e_2}  \right]\, .
\eea
where the last expression holds when $\Im \e_1 > \Im \e_2$  and one can derive a similar one in case $\Im \e_1 < \Im \e_2$ by exchanging $\e_1 \leftrightarrow \e_2$ with an overall minus sign\footnote{These cases probably form different elements of the discrete matrix above and below the diagonal, since the poles of the inverted oscillator dos are at $\Im \e_1 = n_1 + \half$. The matrix is then appropriately real and antisymmetric}. We expect that our analytically continued result is valid for some specific branch. A different branch would give a different normalization. This difference in normalization we expect to play a role in the contribution of non-perturbative states as discussed in the main text.
The result can also be written as
 \bea\label{finalkernel}
O(\e_1, \e_2) &=&\frac{2 e^{(\e_1 + \e_2)\pi/2} }{(1+e^{2 \pi \e_2})^{1/4}(1+e^{2 \pi \e_1})^{1/4}} \bigg| \frac{\Gamma(1/4+ i \e_1/2)}{\Gamma(3/4+ i \e_1/2)} \frac{\Gamma(3/4+ i \e_2/2)}{\Gamma(1/4+ i \e_2/2)} \bigg|^\half   \frac{1-e^{(\e_2 - \e_1)\pi/2}}{\e_1 - \e_2} \, \nn \\
&=&  \frac{1}{\pi} \lvert \Gamma(1/4+ i \e_1/2) \Gamma(3/4+ i \e_2/2)\rvert e^{\pi (3 \e_2 + \e_1)/4}   \frac{  \sinh\left(\frac{1}{4} \pi
   (\e_2-\e_1)\right)}{\e_1-\e_2}  \, . \nn \\
\eea

\subsubsection{Calculation of the integrals for segments}
We will now perform a consistency check and understand better our branch choice. We split the integrals into sections with respect to the branch points. We demand that the parameter $t$ is real and we drop the regulator. Then we indeed find a result that differs for different sections of $t$. The sections are $(- \infty , -1/2)  \cup (-1/2, 1/2)  \cup (1/2, \infty)$ . We have computed the integrals for each section by taking the limit at the branch points sending a small parameter to zero (ex. we integrate up to $1/2+\epsilon$ and then we send $\epsilon\rightarrow 0$). The results are (to be multiplied with the normalization prefactors $N(\epsilon_{1}), \,N(\epsilon_{2})$)
\begin{itemize}
\item For the section $(0 , 1/2) $: 
\be\label{ip1}
I^{+}(t)=\frac{\sqrt{2 \pi } e^{\frac{\pi  {\epsilon}_{1} }{2}} \left(\frac{2}{2 t+1}-1\right)^{+ \half i \epsilon_1}}{(1-4 t^2)^{1/4}}
\ee
\be\label{im1} 
I^{-}(t)=\frac{\sqrt{2 \pi } e^{\frac{\pi  {\epsilon}_{2} }{2}} \left(\frac{2}{2 t+1}-1\right)^{-\frac{1}{2} i{\epsilon}_{2}   }}{\left(1-4 t^2\right)^{3/4}}
\ee
\item For the section $(1/2 , \infty) $: 
\be \label{ip2}
I^{+}(t)=\frac{(1-i) \sqrt{\pi } \left(\frac{2 t-1}{2 t+1}\right)^{\frac{i {\epsilon}_{1}}{2}}}{(4 t^2-1)^{1/4}}
\ee

\be \label{im2}
I^{-}(t)=-\frac{(1-i) \sqrt{\pi } \left(\frac{2 t-1}{2 t+1}\right)^{-\frac{1}{2} (i {\epsilon}_{2} )}}{\left(4 t^2-1\right)^{3/4}}
\ee

One can similarly obtain the rest of the sections by $t \rightarrow - t$.
\end{itemize}
One can now notice that \ref{ip1},~\ref{ip2} are the same expression if one chooses $-1=e^{-i\pi}$ and \ref{im1},~\ref{im2} are the same if we chose $-1=e^{i\pi}$. This choice corresponds to picking a specific branch. We already know that the spectrum of the inverted oscillator is twofold degenerate and our choice just means that the even/odd modes live in a different sheet of the complex energy plane. After changing variables $z=1/\half+t$ this choice gives the same integral and result as in~\ref{hypergeometricresult}

\subsubsection{The sine/sinh kernel}

It is now easy to see that since $O(\e_1, \e_2)= A(\e_1) K^{sinh}(\e_1- \e_2) B(\e_2) $, the only interesting asymptotic contribution comes from the kernel in the middle. The diagonal normalization factors can be shown to contribute non-perturbatively, since they do not admit an $1/\e$ expansion and scale for large $\e$ as $e^{a \e}$ with $a$ a parameter depending on the branch we choose. The kernel whose spectrum we want to compute is the analytic continuation of the very well studied sine kernel  $K^{sine}(\e_1, \e_2) = \frac{\sin(\frac{1}{4} \pi (\e_1 - \e_2)) }{\e_1 - \e_2}$ for which various results exist in the literature in relation to its spectrum and Fredholm determinants~\cite{Dyson:1976nq,Deift,Krasovsky}. 
\\
\\
One way of computing its determinant is to discretise and bring it into a Toeplitz form. In our case one can put a cutoff $\Lambda$ and then use the density of states of the inverted oscillator $i \e_j = j+\half$ which is equidistant, or equivalently analytically continue in $\omega$. Then calculating the determinant of the sine kernel with support on an energy segment one finds that it can be represented as a Toeplitz determinant in a scaling limit
\bea
\det K^{sine}|_{-\infty}^{-\mu} &=& \det\left(1- K^{sine}|_{-\mu}^0 \right)\, , \quad 
\Leftrightarrow \lim_{N \rightarrow \infty} N \det C_{j - k}\, , \nn \\  \text{with} \quad C_{j - k} &=& \delta_{j k} - \frac{\sin(\frac{\pi \mu}{2 N} (j - k))}{\pi (j-k)}
\eea
We will now discuss some properties of this fredholm determinant, and provide an asymptotic evaluation for large $\mu$, with which we can match the torus contribution to the twisted states.

\subsubsection{Level spacings}\lab{LevelSpa}

The level spacing distribution $E_\beta(n, \mu)$ of Random matrices is the probability that the interval  $(0, \mu)$ contains exactly n eigenvalues~\cite{Mehta}. In our case these will be energy eigenvalues and the random matrix is the Hamiltonian. Thus the Hilbert transform operator effectively randomizes the energy eigenvalues of the system which are to be drawn from an ensemble  (GUE/GOE/GSE). The parameter $\beta$ denotes the ensemble and for us $\beta=2$ (GUE). We first define $D (\mu ; \lambda) = \det \left(1-\lambda K^{sine} \right)$. We also define  $K_\pm = K^{sine}(x, y) \pm K^{sine}(x, - y)$ and similarly $D_\pm (\mu ; \lambda) = \det \left(1-\lambda K^\pm \right)$. For the other ensembles, $\beta$ , the kernel is a matrix. Then one has~\cite{Mehta}
\be
E_2(n; (0,\mu)) = \frac{(-1)^n}{n!} \frac{\partial D(\mu; \lambda) }{\partial \lambda^n} |_{\lambda = 1}
\ee
and for the other ensembles one can again find formulas involving $E_\pm$ $D_\pm$. We will now use the asymptotic formulas in the literature for the level spacings as $\mu \rightarrow \infty$ much like what we want for the asymptotic expansion of string theory $\mu \rightarrow \infty$. We provide here the more general result/conjecture for arbitrary $\beta, n$~\cite{Forrester} that correctly reproduces the proven result for $n=0, \beta=2$~\cite{Dyson:1976nq,Deift,Krasovsky}
\bea
\log E_\beta (n ; (0,\mu)) \sim_{\mu \rightarrow \infty} - \beta \frac{\mu^2}{16} + \left(\beta n + \beta/2 -1 \right) \frac{\mu}{2} \nn  \\
+ \left[ \frac{n}{2}\left(1 - \beta/2 - \beta n/2 \right) + \frac{1}{4} \left(\beta/2 + 2/\beta - 3 \right) \right]  \log \mu  +....
\eea
We now need to remember that the Pfaffian is the square root of the determinant and that we need to divide our result by an extra factor of 2, since we want to match the bosonic string theory partition function, that has support on the one side of the potential. After taking these into account, one finds the twisted state contribution
\be 
\Theta =\frac{1}{4} \log E_2 (0 ; (0,\mu)) =  - \frac{1}{32} \mu^2 - \frac{1}{16} \log \mu +\frac{1}{48}\log 2 + \frac{3}{4}\zeta'(-1) + O\left( \frac{1}{\mu^{2m}}\right) \, . 
\ee
We see that we correctly capture only closed string contributions with even higher powers of $1/\mu$ and some of these coefficients can be found in~\cite{Dyson:1976nq}. Moreover this formula predicts that there is no-logarithmic divergence coming from the genus $0$ spherical contribution.  As a bonus, it is interesting to note that one can make the same computation with orthogonal or symplectic matrices in $GOE, GSE$ which can be found to receive open string corrections with odd powers in $\mu$. These results might be relevant for the unoriented string theory on the orbifold, where odd powers of $\mu$ are known to appear and orthogonal/symplectic symmetries to be relevant.

\subsubsection{Properties of the sine kernel}

The sine kernel has some remarkable properties some of which which we list here 
\begin{itemize}
\item Its eigenfunctions are the prolate spheroidal functions and some asymptotic forms of the spectrum exist. 
\item  The Christoffel Darboux (CD) kernels approach the sine kernel in a scaling limit that focuses on the bulk of the spectrum. 
\item As with all the CD kernels it is a self-reproducing kernel, it obeys $K*K = K$. 
\item It is the band-limited version of the Dirac delta distribution. To understand this better, let $f \in L^2(\mathbb{R})$ a function whose fourier transform has support on the segment $[- \pi b, \pi b]$ (band limited functions) Then the sine kernel is an orthogonal projection to this space since
\be 
\int_{-\infty}^\infty d y \frac{\sin(\pi b (x-y))}{\pi(x-y)} f(y)  \, = \frac{1}{\sqrt{2 \pi}} \int_{- \pi b}^{\pi b} e^{i x \xi} \mathcal{F}[f](\xi) d \xi
\ee
\item Moreover one can further consider functions $f \in L^2([-s, s])$. This gives both energy and time band limited functions (in our case $s\sim \mu$ is the energy-band limit while $b=1/4$ is a ``time-band'' limiting). This is called a compression of the sine kernel and gives a trace class operator.
\item It is easy to see that it is the natural regulating description of the dirac-$\delta$ function we were expecting to have (for $O^2$), since at the discrete level we encountered the identity operator $\delta_{n m}$ and we were filling eigenvalues up to the size of the matrix $N$. It also allows for a rigorous understanding of limiting the energy and defining the fermi surface which corresponds to filling all the negative energy states up to a band below $0$ corresponding to the chemical potential $-\mu$.
\end{itemize}





\end{document}